\documentclass[usenatbib]{mnras}

\usepackage{graphicx}	
\usepackage{amsmath,amssymb,amsfonts,mathrsfs} 
\usepackage{bm}		
\usepackage{soul}
\usepackage[flushleft]{threeparttable}
\usepackage{multirow}
\usepackage{xspace}
\usepackage{Custom_Functions}

\defcitealias{gaikwad2016}{Paper-I}	
\defcitealias{puchwein2015}{P15}		
\defcitealias{haardt2012}{HM12}		
\defcitealias{khaire2015a}{KS15}		
\defcitealias{khaire2018a}{KS18}		
\defcitealias{gaikwad2018}{G18}	

\usepackage[usenames, dvipsnames]{color}
\definecolor{mycolor}{RGB}{0,128,0}
\definecolor{newaddcolor}{RGB}{255,0,255}

\usepackage[normalem]{ulem}

\title[Transmission spikes in the \lya forest]{Probing the thermal state of the intergalactic 
medium at $z>5$ with the transmission spikes in high-resolution 
\lya forest spectra}

\author[Gaikwad et.al]{%
Prakash Gaikwad$^{1}$ \thanks{E-mail: \href{pgaikwad@ast.cam.ac.uk}{pgaikwad@ast.cam.ac.uk}}, 
Michael Rauch$^{2}$, Martin G. Haehnelt$^{1}$, Ewald Puchwein$^{3}$,
\newauthor{James S. Bolton$^{4}$, Laura C. Keating$^{5}$, Girish Kulkarni$^{6}$, Vid Ir\v{s}i\v{c}$^{1}$,}
\newauthor{Eduardo Ba\~nados$^{7}$, George D. Becker$^{8}$, Elisa Boera$^{8,9,10,11}$,Fakhri S. Zahedy$^{2}$, }
\newauthor{Hsiao-Wen Chen$^{12}$, Robert F. Carswell$^{1}$, Jonathan Chardin$^{13}$ and Alberto Rorai$^{1}$}
\\
$^{1}$Institute of Astronomy and Kavli Institute for Cosmology, Cambridge University, Madingley Road, Cambridge CB30HA, UK\\
$^{2}$Carnegie Observatories, 813 Santa Barbara Street, Pasadena, CA 91101, USA\\
$^{3}$Leibniz-Institut f\"ur Astrophysik Potsdam (AIP), An der Sternwarte 16, 14482 Potsdam, Germany\\
$^{4}$School of Physics and Astronomy, University of Nottingham, University Park, Nottingham, NG7 2RD, UK\\
$^{5}$Canadian Institute for Theoretical Astrophysics, 60 St. George Street, University of Toronto, ON M5S 3H8, Canada \\
$^{6}$Department of Theoretical Physics, Tata Institute of Fundamental Research, Homi Bhabha Road, Mumbai 400005, India\\
$^{7}$Max Planck Institut f\"ur Astronomie, K\"onigstuhl 17, D-69117, Heidelberg, Germany\\
$^{8}$Department of Physics \& Astronomy, University of California, Riverside, CA, 92521, USA\\
$^{9}$SISSA, Via Bonomea 265, 34136 Trieste, Italy\\
$^{10}$INAF-Osservatorio Astronomico di Trieste, Via Tiepolo 11, I-34143 Trieste, Italy\\
$^{11}$IFPU, Institute for Fundamental Physics of the Universe, Via Beirut 2, 34014 Trieste, Italy\\
$^{12}$Department of Astronomy \& Astrophysics, The University of Chicago, Chicago, IL 60637, USA\\
$^{13}$Observatoire Astronomique de Strasbourg, Universit ́\'e de Strasbourg, CNRS UMR 7550, 11 rue de l'Universit \'e, F-67000 Strasbourg, France}

\date{}

\pubyear{2015}

\begin{document}
\label{firstpage}
\pagerange{\pageref{firstpage}--\pageref{lastpage}}
\maketitle


\begin{abstract}
We compare a sample of five high-resolution, high S/N  \lya forest spectra of
bright $6<z<\sim 6.5$ QSOs aimed at spectrally resolving the last remaining
transmission spikes at $z>5$ with those obtained from mock absorption spectra
from the \sherwood and \relics suites of hydrodynamical simulations of the
intergalactic medium (IGM). We use a profile fitting procedure for the inverted
transmitted flux, $1-F$, similar to the widely used Voigt profile fitting of
the transmitted flux $F$ at lower redshifts, to characterise the transmission
spikes that probe predominately underdense regions of the IGM.  We are able to
reproduce the width and height distributions of the transmission spikes, both
with optically thin simulations of the post-reionization Universe using a
homogeneous UV background and full radiative transfer simulations of a late
reionization model.  We find that the width of the fitted components of the
    simulated transmission spikes is very sensitive to the instantaneous
temperature of the reionized IGM.  The internal structures of the spikes
    are more prominent in low temperature models of the IGM. The width
distribution of the observed transmission spikes, which require high spectral
resolution ($\leq $ 8 \kmps) to be resolved, is reproduced for optically thin
simulations with a temperature at mean density of $T_0= (11000 \pm
1600,10500\pm 2100,12000 \pm 2200)$ K at $z= (5.4,5.6,5.8)$.  This is weakly
dependent on the slope of the temperature-density relation, which is favored
to be moderately steeper than isothermal.  In the inhomogeneous, late
reionization, full radiative transfer simulations where islands of neutral
hydrogen persist to $z\sim5.3$, the width distribution of the observed
transmission spikes is consistent with the range of $T_0$ caused by spatial
fluctuations in the temperature-density relation.

\end{abstract}



\section{Introduction} \label{sec:introduction}
\vspace{-2mm}

The reionization of intergalactic \HI and \HeI by ultraviolet photons from
stars and black holes in the first galaxies is one of the major phase
transitions of the universe
\citep{fan2001,fan2006,robertson2010,bolton2011,planck2014,planck2018}.
Photo-heating during the reionization increases the temperature of the
intergalactic medium (IGM) \citep{hui1997,trac2008,mcquinn2009,mcquinn2015a}.
The thermal and ionization histories of the Universe are thus interlinked, and
can be used in tandem to understand the process of reionization and the nature
of ionizing sources
\citep{haehnelt1998,furlanetto2009,mcquinn2011t,becker2015a,chardin2017,davies2018,keating2018,kulkarni2019a}.

The \HI \lya forest observed along sightlines towards bright QSOs is
frequently used to constrain the thermal and ionization state of the
IGM at $z < 5$.  For underdense or moderately overdense regions, the
thermal state of the IGM is often approximated as a power law,
parameterized as the temperature at mean density and the slope of the
power law \citep[$T_0$ and
  $\gamma$][]{lidz2010,becker2011,rudie2012,garzilli2012,bolton2014,boera2014,hiss2018,telikova2018,telikova2019,walther2019,boera2019}.
Other relevant parameters are the \HI photo-ionization rate
\citep[\GHI][]{rauch1997,bolton2007,faucher2008c,calverley2011,becker2013,gaikwad2017a,gaikwad2017b,viel2016,khaire2019b}
and the pressure smoothing scale
\citep{gnedin1998,peeples2010,girish2015,lukic2015,rorai2017}. Due to
the rapid increase of the \lya opacity with redshift, the transmitted
\lya flux at $z>5$ is close to zero, with the occurrence of a few
transmission spikes indicating that the reionization process is
inhomogeneous \citep{becker2015a,bosman2018,eilers2018}.  Analysis of
the transmission spikes towards ULAS J1120+0641 QSO \citep[$z_{\rm
    em}=7.084$,][]{becker2015a,barnett2017} with numerical simulations
suggests that these spikes correspond to underdense, highly ionized
regions of gas \citep{gnedin2017,chardin2018b,kakiichi2018,garaldi2019,nasir2019}.
These studies find that the number and height of the spikes are
sensitive to the ionization fraction (\qion{x}{HII}) of the IGM, which
in turn depends on the photo-ionization rate and the temperature of
the gas in the ionized regions. Perhaps somewhat surprisingly,
however, \citet{garaldi2019} found that the spike shape (especially
the widths of the spikes) appeared to be only weakly correlated with
the temperature of the IGM.  We revisit this question here with a
larger sample of higher resolution, higher quality \lya forest spectra
which we compare to high-resolution hydrodynamical simulations of the
IGM using the \sherwood and \relics simulation suites.  These
incorporate simulations with a homogeneous UV background as well as
full radiative transfer simulations of inhomogeneous reionization
\citep[][Puchwein et.al. 2020 in prep]{bolton2017,kulkarni2019a}.
Note that in the main analysis we used the \relics simulation suite,
but complemented these with simulations from the \sherwood simulation suite
for  additional tests performed in the Appendices.

There are several difficulties in probing the effect of the thermal
state of the IGM on \lya transmission spikes.  High-redshift QSOs,
while intrinsically luminous, are nevertheless faint and observed
spectra are normally taken at moderate resolution as e.g. offered by
VLT/X-Shooter, {\it i.e.} with $\rm ~35 $ \kmps or worse.  As a result
most spectra of high-redshift transmission spikes are of rather modest
quality and the number of well observed transmission spikes is
necessarily still small due to the rarity and faintness of high-$z$
background QSOs \citep{kulkarni2019b}.  We improve this situation here
and present a sample of 5 high resolution (${\rm FWHM} \sim 6$ \kmps)
and high \SNR ($\sim 10$) QSO absorption spectra obtained using the
Magellan Inamori Kyocera Echelle (MIKE) spectrograph on the Magellan
II telescope \citep{bernstein2003}, and the High-Resolution Echelle
Spectrograph (HIRES) on the Keck I telescope \citep{vogt1994}.

Similarly, simulated transmission spikes may be drawn from simulations
with moderate mass or spatial resolution that is only sufficient to
produce mock spectra mimicking moderate resolution spectrographs like
X-Shooter \citep[$35 $ \kmps, but see][]{garaldi2019}.  The thermal
smoothing scale of the \lya transmitted flux due to the IGM
temperature ($T \sim 10^4 \: {\rm K}, b \sim 13$ \kmps) is, however,
significantly smaller than this.  As discussed in detail by
\citet{bolton2009}, rather high mass or spatial resolution is required
to resolve the small scale structure in the underdense regions of the
IGM probed by \lya forest spectra at high redshift.  Previous
theoretical work has focused on analyzing the spikes in
radiative transfer simulations
\citep{gnedin2017,chardin2018b,garaldi2019}.  While these simulations
are physically well motivated, they are computationally expensive and
it is hard to disentangle the effect of the thermal state of the IGM on
the transmission spikes from the reionization history and numerical
limitations.

We have thus chosen to analyze the transmission spikes first in very
high resolution, high dynamic range optically thin simulations with
different thermal and reionization histories where a single parameter
is varied at a time while keeping other parameters fixed.  We
therefore use simulations from the \sherwood and \relics simulation
suite \citep[][ Puchwein et.al. 2020 in prep]{bolton2017} to show how
spike properties depend on the IGM thermal state, the \HI
photo-ionization rate \GHI, and (in the appendices) the pressure
smoothing scale, as well as the mass resolution and box size of the simulations. Once we
have established this we investigate the effect of inhomogeneous
reionization in more physically motivated, spatially inhomogeneous \HI
reionization simulations including radiative transfer
\citep[see][]{kulkarni2019a,keating2019}.

Another problem when comparing simulated and observed \lya forest
spectra is the accurate characterisation of the transmission spike
properties. Transmission spikes are often asymmetric and 
transmission features appear ``blended''.  The often used simple
definition of height and width of spikes based on the maximum and FWHM
of the transmitted flux will thus not capture the detailed information
contained in the complex spike shapes, and could be the reason that
the analyses performed so far show little or no correlations with
astrophysical parameters \citep[see e.g. the width vs temperature
  correlation in][]{garaldi2019}.  Characterizing the shape of
transmission spikes becomes even more crucial for high \SNR, high
resolution QSO absorption spectra.  In practice the problem is very
similar to that of the characterisation of \lya absorption lines
at lower redshift.  To utilise the practical experience gained in this
area with existing software packages \citep[e.g.][]{gaikwad2017b} we
characterize the transmission spikes by fitting Voigt profiles to the
``inverted'' transmitted flux, $1-F$. We will later show that the
fitted parameters obtained in this way are well correlated with
physical properties (e.g. the density and temperature) of the gas
associated with the transmission spikes.  We will also study how the
statistics of the fitted parameters depend on the astrophysical
parameters \GHI, $T_0$ and $\gamma$. Unlike for absorption lines,
there is no direct physical motivation for fitting Voigt profiles to 1-F, so
this should be considered as a purely heuristic approach to comparing
simulated and observed spectra. However, as we will see this does not 
mean that the fit parameters obtained do not correlate with physical 
properties.

The main goal of this paper is to constrain the thermal state of the IGM at
$5.3 \leq z \leq 5.9$.  The paper is organized as follows: In \S
\ref{sec:spike-qualitative-analysis} we present the high resolution spectra of
5 $z>6$ QSOs and discuss qualitatively the physical origin of \lya transmission
spikes. We present the properties of transmission spikes in optically thin
simulations in \S \ref{sec:spike-in-OT-simulation-0} and \S
\ref{sec:spikes-in-OT-simulation}.  We demonstrate the sensitivity of spike
statistics to the IGM thermal state in \S \ref{sec:spike-statistics}. The main
results of the paper are presented in \S \ref{sec:result-OT} and \S
\ref{sec:spikes-in-RT-simulation} by comparing the observed spike statistics
with those from optically thin and radiative transfer simulations. We summarize
our findings in \S \ref{sec:conclusion}.  We assume a flat $\Lambda {\rm CDM}$
cosmological model $(\Omega_{\Lambda},\Omega_{m},\Omega_{b},\sigma_8,n_s, h,Y)
\equiv$ $(0.692,0.308,0.0482,0.829,0.961,0.678,0.24)$ consistent with
\citet{planck2014,planck2018}.  All distances are given in comoving units
unless specified.  \GHI expressed in units of $10^{-12} \: {\rm s}^{-1}$ is
denoted by \GTW.

\section{Transmission spikes in high-resolution, high-redshift \lya forest spectra}
\label{sec:spike-qualitative-analysis}

\subsection{Observations: high-resolution spectra of transmission spikes}

The data consist of the high resolution echelle spectra of five
recently discovered $z>6$ QSOs. The objects were chosen for their brightness,
and individual targets were further selected to maximize the exposure time during a given observing run.  Table \ref{tab:observations}
gives the name of each object, the emission redshift z$_{\rm em}$, the J AB band magnitude
\citep[from the compilation of][]{ross2019}, the
total on-source exposure time T in hours, and a typical signal-to-noise ratio per
pixel.  The objects were observed under mostly photometric conditions
in sub-arcsec seeing.  The first four objects were observed with the
MIKE instrument \citep{bernstein2003} on the Magellan II telescope at
Las Campanas Observatory.  A 0.5" wide slit gave a measured spectral
resolution of 5 km s$^{-1}$ (FWHM). The spectra were binned onto 2 km
s$^{-1}$ wide pixels.  The spectrum of SDSSJ010013.02+280225.8 was
obtained with the HIRES instrument \citep{vogt1994} on the Keck I
telescope, and a 0.861" wide slit, giving a resolution of 6.1 km
s$^{-1}$ (FWHM), sampled by 2.5 km s$^{-1}$ wide bins.  The data were
reduced with a custom pipeline \citep{becker2012}. Optimal
sky-subtraction on the individual, un-rectified exposures was
performed according to the prescription by \citet{kelson2003}.

For the continuum model, a power law was assumed. As the high resolution
echelle spectra only had limited coverage of the unabsorbed region redward of
the Ly$\alpha$ emission, the continuum was derived from flux-calibrated, low
resolution spectra of the same QSOs which extended further to the red. For
objects 1-3 (in table \ref{tab:observations}), the continua were determined
from the discovery spectra, using fits to regions redward of Ly$\alpha$
avoiding broad emission lines, whereas the power law slopes for objects 4 and 5
were taken from the literature cited.  The continuum was then scaled uniformly
to match the flux-calibrated high resolution spectrum in the overlap region
with the low-resolution spectrum redward of Ly$\alpha$, and divided into the
spectrum. To correct for the rapidly variable region of the spectrum near the
Ly$\alpha$ emission line, the emission line region was fitted with a higher
order polynomial in the previously continuum divided spectrum, which then was
multiplied into the previous continuum fit. The final continuum thus obtained
was divided into the data.  As we show later, the width of the fitted components
of the transmission spikes are relatively robust to continuum fitting
uncertainty.

During data reduction a certain degree of smoothing is introduced into the
data that appears as a discrepancy between the RMS fluctuations in the final
data and the propagated error array, and generally results in an underestimate
of the reduced $\chi_{\nu}^2$ when fitting line profiles. To counteract this
problem,  a correction factor (generally a number close to unity varying slowly
with wavelength) was derived from the observed ratio between the RMS
fluctuations and the error array, as determined from wavelength windows in the
spectrum with zero flux.  A linear fit to the correction factor as a function
of wavelength was divided into the error array to obtain a reduced
$\chi_{\nu}^2 \sim 1$ during profile fitting.

\begin{table*}
\scriptsize
 \centering
  \caption{Properties of the QSO spectra analysed in this work (see
    text for further details).}
  \begin{tabular}{rcccccccr}
\hline  \hline 
 ID  & Name   & $z_{\rm em}$ & J & ${\rm T} [\rm h]$   & (S/N)/{\rm pixel} & Reference& Instrument& Dates of the observations  \\
 \hline
1 & ATLASJ158.6938-14.4211 & 6.07&19.27    &   11.8  &  7.7  & \citet{chehade2018} & MIKE  & March and April,  2018 \\
2 & PSOJ239.7124-07.4026 & 6.11 & 19.37  &   10.0  &    6.5  & \citet{banados2016} & MIKE  & March, April and June, 2018\\
3 & ATLASJ025.6821-33.4627 & 6.34& 19.10   &   6.7  &    12.0  & \citet{carnall2015} & MIKE &  October and December, 2018\\
4 & J043947.08+163415.7 & 6.51  &17.47  &   10.0  &     30.3  & \citet{fan2018} & MIKE  & October and December, 2018\\
5 & SDSSJ010013.02+280225.8 &6.30  &17.60  &   5.0  &     20.0  & \citet{wu2015} & HIRES & November 2017 \\
\hline \hline
\end{tabular}
\\
\label{tab:observations}
\end{table*}

\subsection{Characterising width and height of individual components}

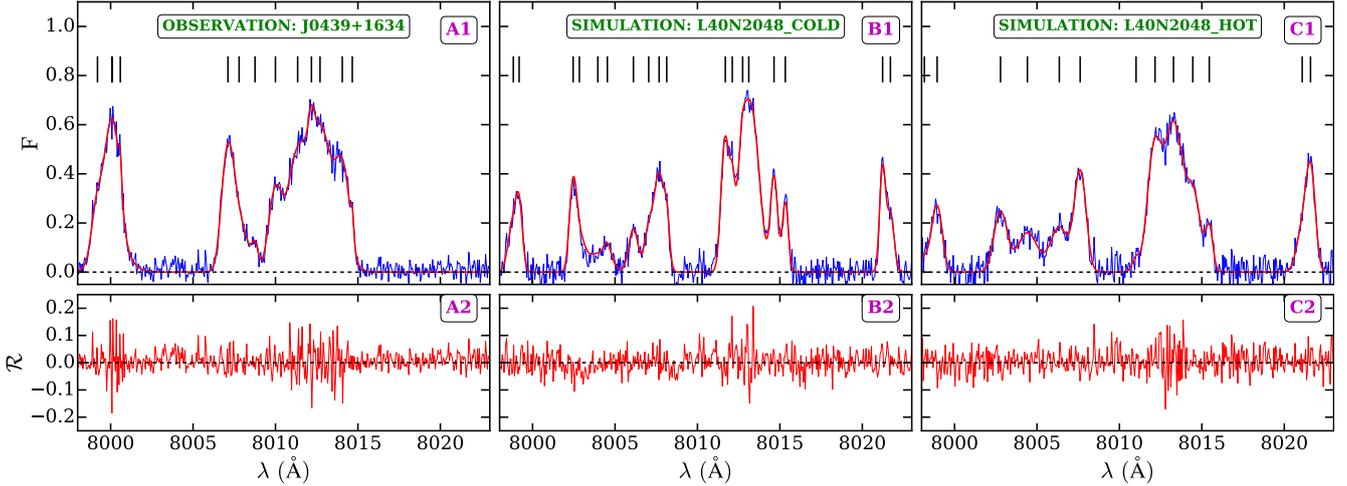
\InputFigCombine{VIPER_Fit_Example-0.pdf}{175}{Examples of
  transmission spikes from observed spectra (panel A1) and simulated
  spectra from \cold (panel B1) and \hot (panel C1) optically thin
  simulations drawn from the \relics simulation suite at $z \sim 5.59$.  
  The widths and  heights of the spikes are sensitive to the thermal and ionization
  state of the IGM.  The shape and number of spikes in the observed
  spectra are similar to the simulated spectra from the \hot model.
  The resolution and noise properties of the simulated spectra are
  chosen to match the observed spectra.  As described in the main text
  we fit the ``inverted'' transmitted flux, $1-F$, with
  multi-component Voigt profiles using the software package \viper
  described in detail in \citet{gaikwad2017b}.  In \viper, the number of 
  components to be fitted in given region is decided automatically by 
  minimizing the Akaike information criteria with correction 
  \citep[see][for details]{gaikwad2017b}.
  The top panels show the input spectra ($F$,
  blue solid curve) and fitted spectra ($F$,in red solid curve). The
  black solid lines mark the location of the centres of Voigt
  components identified and fitted by \viper.  The bottom panels show
  that the residuals between the input and fitted spectra are random
  and less than 11 per cent.  The number of components identified in
  the \cold model is larger than in the \hot model due to the smoother
  transmitted flux distribution.}{\label{fig:viper-fit-example}}

\figref{fig:viper-fit-example} compares transmission
spikes in a high-resolution \lya forest observation of the QSO J043947.08+163415.7 with the MIKE
spectrograph with simulated spikes drawn from the \relics simulation
suite (see Section~\ref{subsec:sherwood_relics}, Puchwein et.al. 2020 in prep), for \cold
and \hot models with a spatially uniform UV background. The
transmission spikes have complex shapes composed of many asymmetric
and blended features.  Even isolated transmission spikes are often
highly asymmetric and consist of two or more ``components''.  In order
to facilitate a more quantitative discussion of the transmission
spikes, we focus on two properties, namely the height and width of
individually identifiable components.  We quantify these (see \S
\ref{sec:spikes-in-OT-simulation}) by fitting the ``inverted''
transmitted flux, $1-F$, with multi-component Voigt profiles, similar
to the fitting of Voigt profiles of the transmitted flux, $F$, often
employed at lower redshift to characterise absorption lines.

The best fits to observed and simulated spectra are shown in
\figref{fig:viper-fit-example} by the red curve.  The location of individual
identified components are marked by black vertical lines.
\figref{fig:viper-fit-example} show simulated spectra for the \hot and \cold
model and  illustrate the sensitivity of spike features to the thermal state of
the IGM.  The main effect of increase in IGM temperature is to reduce the
internal structure of the spikes (i.e., more blending). As a result, fewer
and broader components are required to fit the transmission spikes in the \hot
model. In general, the spikes in the \hot model are (i) broader, (ii)
larger in height, (iii) more asymmetric and (iv) more blended (hence fewer
in number) than those in the corresponding \cold model.  It is interesting to
note that the number and shape of the spikes in the \hot model are
qualitatively very similar to those in the observed spectra. We will analyse
this more quantitatively below.

\subsection{The origins of transmission spikes}
\label{sec:spike-theoretical-analysis}

The complex shapes of the transmission spikes shown in the last
section will be a superposition of features in the line-of-sight
distribution of density, temperature, photo-ionization rate and
peculiar velocity.  To get a better feel for this in
Fig. \ref{fig:spike-origin} we show mock spectra where we isolate the
effect of varying these parameters, one at a time.
Fig. \ref{fig:spike-origin} illustrates how the variation in any of
these physical quantities along a sightline can lead to regions of
smaller \HI optical depth and hence transmission spikes.  Panels
A1-A6, B1-B6 and C1-C6 show the effect of underdensity, enhanced \GHI
and enhanced temperature along a sightline, respectively.  All these
effects can result in a smaller number density of neutral hydrogen,
\qion{\rm n}{HI}, along the sightline.  Panels D1-D6 show that a
diverging peculiar velocity field along the sightline can also produce
transmission spikes.  Hot, ionized underdense regions subjected to
high photo-ionization rates and diverging peculiar velocities will
thus produce the most prominent transmission spikes in high$-z$ QSO
absorption spectra \citep{gnedin2017,chardin2018b,garaldi2019}.  The
transmission spikes shown in \figref{fig:spike-origin} are by
construction isolated, symmetric and simple. As discussed in the
previous section, transmission spikes in both observed spectra and
spectra drawn from cosmological hydrodynamical simulations have
complicated shapes, as generally more than one of the above effects
contribute.

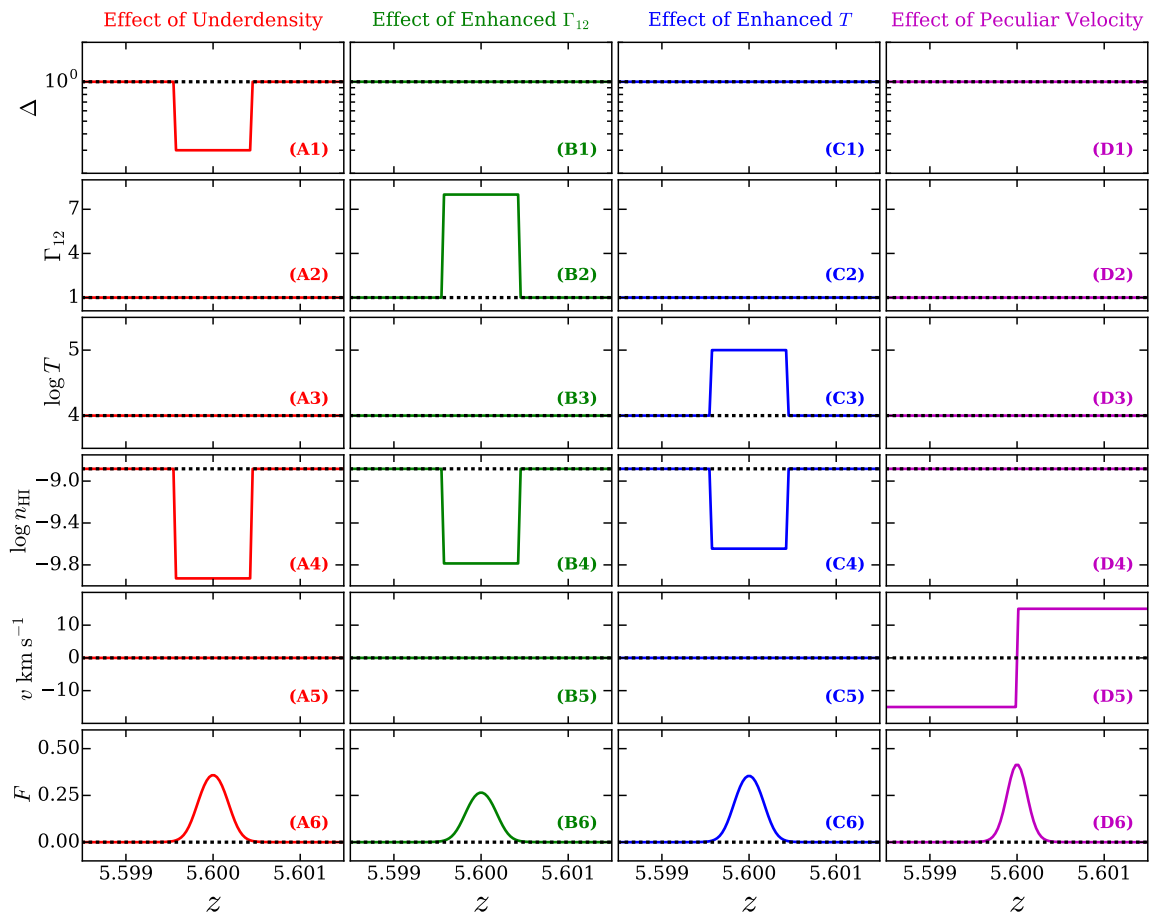
\InputFigCombine{Spike_Origin_Cases.pdf}{150}{Illustration of the
  origin of transmission spikes due to the variation of individual
  physical parameters.  Each row displays the variations in line of
  sight density ($\Delta$), \HI photo-ionization rate ($\Gamma_{\rm
    12}$), temperature ($T$), \HI number density (\qion{\rm n}{HI}),
  peculiar velocity ($v$) and transmitted \lya flux, $F$.  The black
  dashed lines show the default values.  In each column, a different
  physical parameter is varied.  In the first three columns (i.e. for
  varying $\Delta$, \GTW and $T$) the occurrence of spikes is due to a
  change in \qion{\rm n}{HI}.  Panels D1-D6 instead shows the
  formation of a transmission spike due to a diverging velocity flow
  along the sightline while \qion{\rm n}{HI} remains constant.
  Realistic transmission spikes (see \figref{fig:los-comparison}),
  have more complicated shapes and will be due to a combination of
  these effects.  }{\label{fig:spike-origin}}

\section{Transmission spikes in optically thin simulations}
\label{sec:spike-in-OT-simulation-0}
\subsection{The \sherwood and \relics simulation suites}
\label{subsec:sherwood_relics}

We use cosmological hydrodynamical simulations from the \relics
simulation suite to investigate transmission spikes
in the high-redshift \lya forest; see
Table~\ref{tab:simulation-details} for an overview.  
The simulations
were performed with a modified version of the \pgthree code (itself an
updated version of the \gtwo code presented in
\citealt{springel2005}).  The code uses a tree-particle mesh gravity
solver for following cosmic structure formation and a manifestly
energy and entropy-conserving smoothed particle hydrodynamics scheme
\citep{springel2002} for following the hydrodynamics. The \relics
simulations build upon the original \sherwood simulation suite
(which is used in
Appendix~\ref{app:resolution-study} to study numerical convergence) in
that the initial conditions were generated in the same way, and much
of the modeling of the IGM and \lya forest is based on similar
methods \citep{bolton2017}\footnote{\url{https://www.nottingham.ac.uk/astronomy/sherwood/}}.

Our main production runs follow $2 \times 2048^3$ particles in a $(40
\, h^{-1} \, \textrm{cMpc})^3$ volume, corresponding to a gas mass
resolution of $9.97 \times 10^4 \, h^{-1} \, \textrm{M}_\odot$. For
the gravitational softening we adopt $0.78 \, h^{-1} \,
\textrm{ckpc}$.  Star formation is treated with a rather simplistic
but numerically efficient scheme in which all gas particles with
densities larger than 1000 times the mean cosmic baryon density and
temperatures smaller than $10^5$ K are converted to collisionless star
particles. While this does not produce realistic galaxies, it allows
robust predictions of the properties of the IGM
\citep{viel2004a}. Photo-heating and photo-ionization are followed based
on external UV background models.  In a departure from the 
\sherwood simulations, we use a non-equilibrium ionization and
cooling/heating solver \citep{puchwein2015,gaikwad2019} for following the
thermochemistry of hydrogen and helium. This ensures that no
artificial delay between the reionization of gas and its photo-heating
is present.  We have also replaced the slightly modified
\cite{haardt2012} UV background used in \sherwood with the
\textit{fiducial} UV background model from \cite{puchwein2019} in our
\default run. This results in a more realistic reionization history
with hydrogen reionization finishing at $z \approx 6.2$.  The
\cold/\hot models were obtained by decreasing/increasing the \HI and
\HeI photo-heating rates in the \textit{fiducial} UV background model
by a factor of 2 and the \HeII photo-heating rate by a factor of $1.7$,
while keeping all photo-ionization rates fixed. Mock \lya forest
spectra were extracted from all simulations as described in
\citet{bolton2017} \citep[see also][]{gaikwad2018,gaikwad2019}.
Through out this work, we use simulations from the \relics simulation suite
for the main analysis.  Additional simulations from the \sherwood simulation
suite are used for convergence tests as described in Appendix
\ref{app:resolution-study}.

\subsection{The temperature density relation (TDR) in optically thin simulations}
\label{subsec:tdr-cold-hot}
 \InputFigCombine{TDR_L40N2048_COLD_HOT_ATON_PATCHY.pdf}{180}{
   Comparison of the temperature-density relation (TDR) in the \cold (panel A),
   \hot (panel B), \aton (panel C) and \patchy (panel D) simulations at
   $z=5.8$.  The \cold and \hot models correspond to the optically thin \relics
   simulations.  The \aton simulation is post-processed with the radiative
   transfer code ATON, while the \patchy simulation includes the effect of
   pressure/Jeans smoothing as well as the shock heating of the gas.  For
   optically thin simulations the TDR in underdense and moderately overdense
   regions can be approximated as a power-law relation ($T= T_0 \:
   \Delta^{\gamma-1}$) at $\Delta \leq 10$.  The best-fitting relation in the
   \cold and \hot models is shown by the black dashed lines.  By construction,
   the temperature of the gas with $\Delta \leq 10$ in the \hot model is
   consistently larger than that in the \cold model.  As a result, the heights
   and widths of the components to be fitted to the transmission features are
   expected to be different in the \hot and \cold models.  Unlike the optically
   thin simulations, the radiative transfer simulations (\aton and \patchy) do
   not exhibit a single power-law TDR at $\Delta \leq 10$.  For visual
   purposes, we show two power-laws that cover the range in temperature for the
   radiative transfer runs (panel C and D).  The absence of gas with $T>30000$
   K in \aton is because the shock heating is not captured self-consistently in
   the \aton runs.  We plot mass weighted (SPH particle) temperature and
       density in panel A, B and D, while we plot volume weighted temperature
       and density (calculated on grids) for the \aton model (panel C). As a result, the number of
       gas elements (at $\Delta \leq 10$) between the two straight power law
       TDR lines is larger in the \aton simulations than in the \patchy
       simulations.  Note that gas with $\Delta >
   10^3$ and $T<10^5$ K has been converted into stars in all these simulations
   \citep{viel2004a}. We discuss the TDR for optically thin and radiative
   transfer simulations in \S \ref{subsec:tdr-cold-hot} and \S
   \ref{subsec:tdr-aton-patchy},
   respectively. }{\label{fig:tdr-cold-hot-aton-patchy}}

\begin{table*}
\centering
\caption{Origin of transmission spikes due to variation in physical
  parameters in our simulations at $5.5 < z < 5.7$ (as shown in
  Fig. \ref{fig:spike-origin}).}
\begin{threeparttable}
\begin{tabular}{ccccc}
\hline \hline
& \multicolumn{4}{c}{Fraction of spikes (in per cent) showing the effect of} \vspace{2mm}\\
Simulation & Underdensity\tnote{[a]} \hspace{5mm}& Enhanced \GHI\tnote{[b]}\hspace{5mm} & Enhanced $T$\tnote{[c]}\hspace{5mm} & Peculiar velocity\tnote{[d]} \\ 
\hline
\simboxflag{40}{2048}{DEFAULT}\tnote{[e]} & 89.1  & $-$  & 5.7  & 18.8 \\
\simboxflag{40}{2048}{COLD}\tnote{[e]}    & 89.3  & $-$  & 4.9  & 14.8 \\
\simboxflag{40}{2048}{HOT}\tnote{[e]}     & 89.8  & $-$  & 5.5  & 19.9 \\
\simboxflag{40}{2048}{ATON}               & 84.4  & 56.8 & 50.1 & 19.5 \\ 
\simboxflag{40}{2048}{PATCHY}             & 79.2  & 54.7 & 40.3 & 18.4   \\ 
\hline \hline
\end{tabular}
\begin{tablenotes}
\item[a] Fraction of all spikes with \qtau{\Delta} $\leq 1$.
\item[b] Fraction of all spikes with \GHI $\geq \Gamma_{\rm HI,median}$ 
         where $\Gamma_{\rm HI,median}$ is the optical depth weighted median \GHI 
         calculated from all the sightlines (i.e., regions with and without spikes).
\item[c] Fraction of all spikes with \qtau{T} $\geq T_{\rm \tau,median}$ 
         where $T_{\rm \tau,median}$ is the  optical depth weighted median temperature
         calculated from all the sightlines (i.e., regions with and without spikes).
\item[d] Fraction of all spikes with $\Delta v \geq 6$ \kmps where $\Delta v$ is 
    difference between mean velocity on redward and blueward side of spike center. 
    For diverging velocity flow $\Delta v > 0$, whereas for converging velocity 
    flow $\Delta v < 0$. The limit of $\Delta v = 6$ \kmps corresponds to the spectral 
    resolution of the instrument.
\item[e] \simboxflag{40}{2048}{DEFAULT} \simboxflag{40}{2048}{COLD} and \simboxflag{40}{2048}{HOT} are optically 
    thin simulations that do not include fluctuations in \GHI.
\end{tablenotes}
\end{threeparttable}
\label{tab:simulation-details}
\end{table*}

\figref{fig:tdr-cold-hot-aton-patchy} shows the TDR in the \hot and \cold
models of the \relics simulation suite.  We shall compare the \hot and \cold
models to study the effect of temperature on transmission spikes.  Note that
the \hot model is not only hotter than the \cold and the \default model but
also has a flatter temperature density relation,  and that the ionization state
of the gas is also different in the models.  This is because the recombination
coefficient is temperature dependent.  For the same photo-ionization rate the
\HI fraction is therefore smaller in the \hot model.
The models \aton and \patchy incorporate the effect of inhomogeneous UV background
that we describe in \ref{sec:spikes-in-RT-simulation}.

\subsection{Examples of transmission spikes in the hot and cold optically thin simulations}

\figref{fig:los-comparison} shows the relevant physical properties
along the two sightlines in the \hot and \cold models.  
As expected, the \hot model shows (i) a
smoother density ($\Delta$) field in real space, (ii) a
smaller \HI fraction (\qion{x}{HI}), (iii) a larger
temperature ($T$) and (iv) a smoother velocity ($v$) 
field compared to the corresponding \cold model.  The
smoothing of the real-space density field and the velocity field can
be attributed to the increased effect of pressure smoothing in the
\hot model.  The resultant \lya optical depth and transmitted flux
calculated from the $\Delta$, \qion{x}{HI}, $T$ and $v$ fields are
shown in panels A6-B6 and panels A7-B7, respectively.  Note that the
location of spikes in the green and yellow shaded regions in redshift
space differs in real space due to the effect of peculiar velocities.
The transmission spikes in the \hot and \cold models have complicated
shapes, qualitatively similar to that in the observed spectra (\figref{fig:viper-fit-example}).  
The smoother transmission
features in the simulated spectra of the hot model are more similar to
those in the observed spectra than those in the more ``spiky'' spectra
in the \cold model.  

In both models, the transmission spikes correspond to regions of low
\HI optical depth (\qion{\tau}{HI} $\leq 4$, black dashed line).  
The peaks in transmission are well correlated with
those in the optical depth weighted overdensities (\qtau{\Delta} $<
0.8$).\footnote{\qtau{\Delta} accounts for the
  redshift space effect of peculiar velocity on $\Delta$. 
  To assign a single overdensity/temperature to each transmission spike, we calculate 
  flux weighted overdensity/temperature \citep[also see][]{garaldi2019}.}  It is
clear from \figref{fig:los-comparison} that the spikes in the \hot model are
smoother, broader and more prominent than in the \cold model.
Furthermore the number of individual transmission components is
significantly smaller in the \hot model than in the \cold model due to
the ``thermal blending'' of transmission features.  The shape and
number of transmission spikes in our simulated spectra are clearly
sensitive to the thermal and ionization state of the IGM in a manner
that we will quantitatively discuss below.

Table \ref{tab:simulation-details} shows the physical effects
responsible for the occurrence of transmission spikes in the optically
thin \hot and \cold simulations. Most of the spikes ($\sim 89$
percent) in the \hot and \cold models occur in underdense
regions. Around $15$ percent of spikes show a diverging velocity field
along the sightline. The effect of enhanced temperature on the
occurrence of transmission spikes is marginal in both \hot and \cold
optically thin simulations.

In summary, \figref{fig:los-comparison} and Table
\ref{tab:simulation-details} show that underdense (\qtau{\Delta}
$<0.8$), more highly ionized (minimum in \qion{x}{HI}) and hotter
regions along a sightline produce more prominent spikes.  This
motivates us to quantify the shape of spikes and to introduce
statistics that are sensitive to the thermal and ionization parameters
of the IGM.

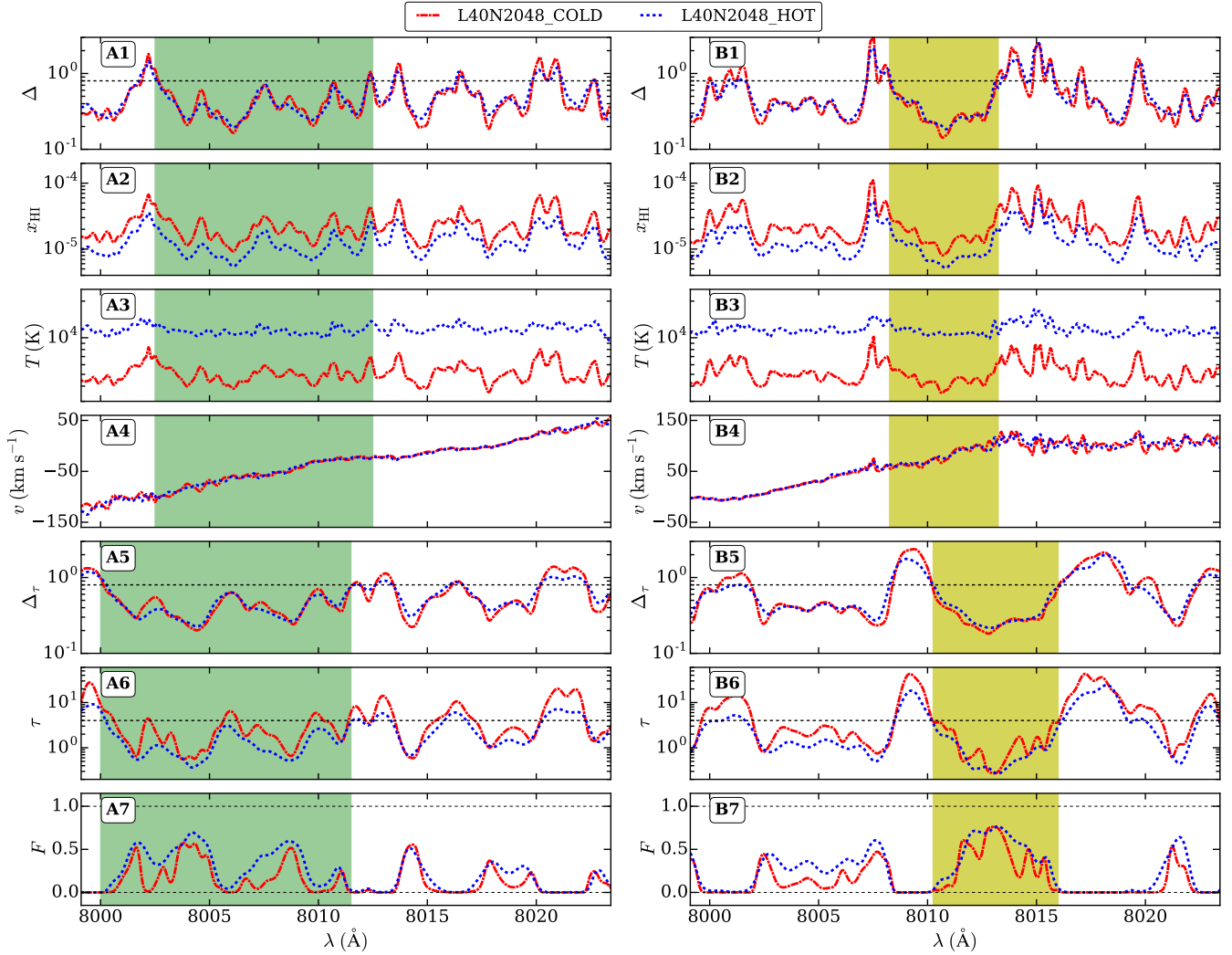
\InputFigCombine{LOS_Comparison.pdf}{175}{Examples of transmission
  spikes in optically thin simulations showing the complex structure
  of the spikes and the dependence on the thermal state of IGM.  The
  figure shows a line of sight comparison of overdensity ($\Delta$,
  panel A1), \HI fraction (\xHI, A2), temperature ($T$, A3), peculiar
  velocity ($v$, A4), optical depth weighted overdensity ($\Delta_{\rm
    \tau}$, A5), \HI \lya optical depth ($\tau$, A6) and transmitted
  \lya flux ($F$, A7) for the \cold (red dot-dashed curve) and the
  \hot (blue dotted curve) optically thin simulations. Panel B1-B7 are
  the same as panel A1-A7 along a different line of sight.  The shaded
  region in panels A7 and B7 show the complex shapes of the spikes in
  regions where the optical depth is lowest ($\tau_{\rm HI} \leq 4$,
  black dashed line in panels A6 and B6).  The shapes
  of the spikes are smoother in the \hot model due to the larger
  temperature (Panel A3 and B3) and smoother density field (panel A1
  and B1) than those in the \cold model.  As a result, the number of
  components identified by \viper is smaller in the \hot model.  The
  transmission spikes also reach larger fluxes in the \hot model due
  to the dependence of the \HI fraction (panel A2 and B2) on
  temperature (via the recombination rate).  The shift of the shaded
  region in panels A1-A4 (B1-B4) as compared A5-A7 shows the effect of
  peculiar velocity on the transmission spikes.  The spikes in the
  simulated spectra occur due to a combination of the effects of
  underdensity, temperature enhancement and peculiar velocity as shown
  in \figref{fig:spike-origin} (\GTW is uniform in the optically thin
  simulations). The mean transmitted flux has not been rescaled for \hot and \cold
  model in the above examples.}{\label{fig:los-comparison}}

\section{Characterising  the properties of  transmission spikes in optically thin simulations}
\label{sec:spikes-in-OT-simulation}
\subsection{Voigt profile fitting of the inverted flux $1-F$}

The shape of absorption features is usually characterized by Voigt
profiles defined by three parameters: (i) the centre of absorption
lines ($\lambda_{\rm c}$), (ii) the \HI column density ($N_{\rm HI}$)
and (iii) the width of the absorption ($b$) features.  Most of the
absorption features in the high-$z$ $(z \sim 6)$ \lya forest are
saturated ($F \sim 0$) and strongly blended. It is well known that
Voigt profile decompositions are highly degenerate for saturated lines
and very sensitive to systematic errors due to continuum fitting and
treatment of noise properties \citep{webb1991,fernandez1996}.  The
inverted transmitted flux, $1-F$, however, becomes similar in
appearance to the absorption features in the \lya forest at lower
redshift, where Voigt profile fitting is much less problematic.  For
convenience, we have thus fitted Voigt profiles to $1-F$, building on
existing experience with Voigt profile decomposition of complex
blended spectral profiles.  At the redshift we consider here the
transmission spikes are observed to be unsaturated, so effectively we use our
\viperfullform \citep[\viper,][]{gaikwad2017b} to fit multi-component
Voigt profiles to the transmission profiles.  Similar to absorption
lines, a simple, isolated and symmetric spike is fitted by $3$
parameters: (i) a spike centre ($\lambda_{\rm c}$), (ii) the logarithm
of the pseudo-column density (denoted by \logaNHI) and (iii) a spike
width ($b$). The pseudo-column density is thereby a measure of the
deficiency of \HI along the sightline where the spike occurs.  For
example, a larger value of \logaNHI means a large \HI deficit hence a
more prominent spike. Our measured \logaNHI are sensitive to the \HI
photo-ionization rate \GHI.  The interpretation of the other two
parameters i.e., $\lambda_{\rm c}$ and width of the spikes remains
unchanged when we fit $1-F$.  As we show below, the distribution of
spike widths is sensitive to the thermal state of the IGM and can be
used to constrain the temperature of the IGM.\footnote{Unlike for
  absorption lines, there is no direct relation between temperature of
  the absorbing gas and the width of the spikes, such that $b_{\rm
    spike} \neq \sqrt{2k_{\rm B}T/m}$.}

Our main aim is to use the distribution of spike widths to constrain
the thermal state of the IGM.  As we will show later, these are much
less sensitive to IGM ionization state and continuum fitting
uncertainties than the heights of the spikes. Note that we
rescale the optical depth in different models to match observations as 
to account for the uncertainty in \GHI at $5.3<z<5.9$ (rescaling
is not applied in \figref{fig:los-comparison}).
We show such rescaling does not significantly affect the widths of the line 
in Appendix \ref{app:observational-systematics}. 

\subsection{The physical properties of the gas probed by transmission spikes}
\subsubsection{The dependence of spike height on density:  \Dtau vs \logaNHI}
\label{subsec:delta-NHI-corr-cold-hot}

\InputFigCombine{log_Delta_vs_log_NHI_L40N2048_COLD_HOT_ATON_PATCHY.pdf}{180}{
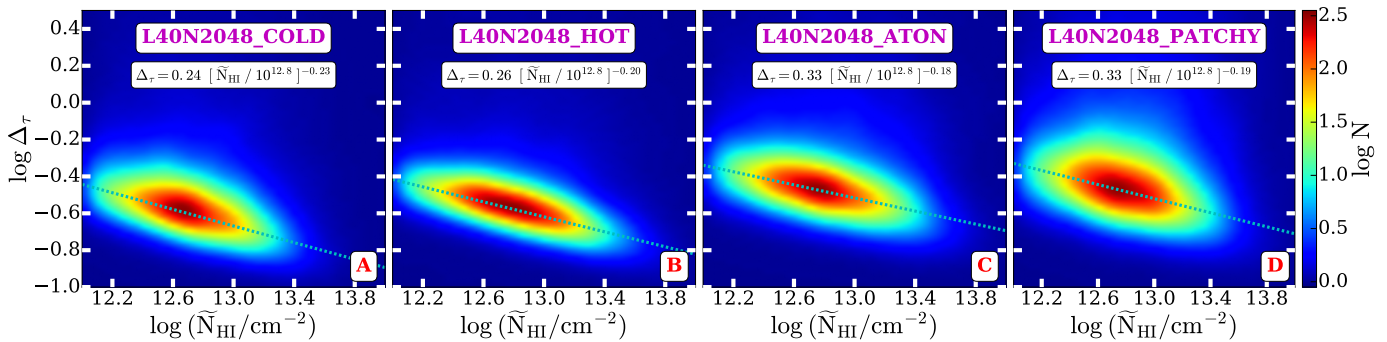
  Panels A, B, C and D show the correlation of optical depth weighted
  overdensity (\qtau{\Delta}) with the pseudo-column density
  (\logaNHI) of transmission spikes in, respectively, the \cold, \hot,
  \aton and \patchy models at $5.5 < z < 5.7$.  Irrespective of the
  model, the spikes correspond to underdensities i.e., $\Delta \sim
  0.25$ for \logaNHI$=12.8$ in the optically thin simulations and
  $\Delta \sim 0.33$ for \logaNHI$=12.8$ in the radiative transfer
  simulations.  The correlation can be fitted with a straight line
  (cyan dotted line) and shows good agreement between the various
  models.  The transmission spikes in the radiative transfer
  simulations are produced by regions with slightly larger densities
  ($\Delta \sim 0.33$) than those in the optically thin simulations
  ($\Delta \sim 0.25$). This is due to the spatial fluctuations in the
  photo-ionization rate and temperature that are present in the
  radiative transfer simulations.  We discuss the \qtau{\Delta} vs
  \logaNHI correlation for the optically thin and radiative transfer
  simulations in \S \ref{subsec:delta-NHI-corr-cold-hot} and \S
  \ref{subsec:delta-NHI-corr-aton-patchy}, respectively.
}{\label{fig:delta-NHI-corr-cold-hot-aton-patchy}}

We now turn to connecting the observed spike shapes to optical depth
weighted physical properties of the IGM in the optically thin
simulations \citep[see Eq. 12 in][]{gaikwad2017a}.  In
\figref{fig:delta-NHI-corr-cold-hot-aton-patchy}, we show the optical
depth weighted overdensity against spike height as measured by
(pseudo) column density \logaNHI for the \cold and \hot
models.  In both models the corresponding overdensity of
spikes is \qtau{\Delta}$<1$ i.e., most of the spikes occur in underdense 
regions.  Note that applying a rescaling of optical depth to
correct for the uncertain amplitude of the UV background results in a
systematic increase or decrease of \logaNHI. However, the
overdensities corresponding to spikes are relatively robust. This is
also evident from \figref{fig:los-comparison}, where the
spikes in the \hot and \cold models correspond to similar
overdensities (rescaling is not applied in
\figref{fig:los-comparison}).

\figref{fig:delta-NHI-corr-cold-hot-aton-patchy} also illustrates that
\logaNHI and \qtau{\Delta} are anti-correlated i.e.  a larger spike
height and higher \HI pseudo-column density corresponds to smaller
overdensity \qtau{\Delta}.  We quantify the degree of anti-correlation
by fitting a straight line of the form ${\rm \Delta_{\tau}} = \Delta_0
\; [{\rm \widetilde{N}_{HI}} \; / \;10^{12.8}]^{\beta}$.  The
normalisation ($\Delta_0=$0.24 for \cold and 0.26 for \hot) and slope
($\beta=$ $-0.23$ for \cold and $-0.20$ for \hot) of the correlation
in the two models are in good agreement with each other. This suggests
that the thermal state of the gas does not have a strong effect on the
\qtau{\Delta}$-$\logaNHI correlation when optical depths are rescaled
to match observations. Note, however, that the scatter in
\qtau{\Delta} for a given \logaNHI (e.g., at \logaNHI $\sim 12.6$)
correlation is smaller in the \hot model. This is because the spikes
are smoother and hence there is less variation in \qtau{\Delta}.

\subsubsection{The dependence of spike width on temperature: \Ttau vs $\log b$}
\label{subsec:T-b-corr-cold-hot}

\InputFigCombine{log_T_gas_vs_log_b_L40N2048_COLD_HOT_ATON_PATCHY.pdf}{180}{
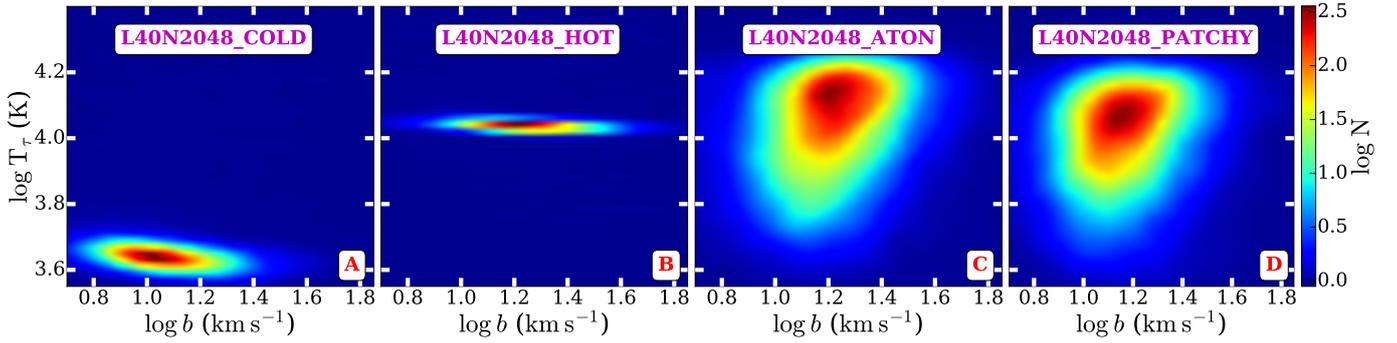
  Panels A, B, C and D show the correlation of optical depth weighted
  temperature (\Ttau) with the line-width parameter ($\log \: b$) of
  spikes in, respectively, the \cold, \hot, \aton and \patchy models
  at $5.5 < z < 5.7$.  The $b$ parameter and \Ttau are systematically
  larger in the \hot model compared to the \cold model.  As shown in
  \figref{fig:tdr-cold-hot-aton-patchy}, \qtau{T} is larger in the
  \aton model compared to the \patchy model due to the presence of
  more gas at $\Delta < 1$ and $T<30000$ K in the \aton model.  The
  scatter in temperature in the RT simulation is much larger than that
  in optically thin simulations due to the presence of UV background
  and temperature fluctuations.  We discuss the \Ttau vs ($\log \: b$)
  correlation for optically thin and radiative transfer simulations in
  \S \ref{subsec:T-b-corr-cold-hot} and \S
  \ref{subsec:T-b-corr-aton-patchy} respectively.
}{\label{fig:T-b-corr-cold-hot-aton-patchy}}

The widths of the components fitted to the spikes are sensitive to the
instantaneous temperature along the sightline (see Fig.
\ref{fig:viper-fit-example} and \figref{fig:los-comparison}).
\figref{fig:T-b-corr-cold-hot-aton-patchy} shows the correlation of optical
depth weighted temperature (\qtau{T}) with spike width ($b$) for the \cold and
\hot model.  The range in temperature associated with spikes is small for both
models.  This is expected as the slope of the TDR
(\figref{fig:tdr-cold-hot-aton-patchy}) for both models is relatively flat and
the temperature associated with $\Delta < 1$ is relatively constant.

\figref{fig:T-b-corr-cold-hot-aton-patchy} illustrates that the spike
widths are well correlated with temperature. The spike widths are
systematically larger in the \hot model ($\log b \sim 1.3$) compared
to the \cold model ($\log b \sim 1.05$).  Even though the range in
temperature is small ($\delta \log T_{\tau}\sim 0.1$) for both models, the
scatter in $\log \: b$ is relatively large ($\delta \log \: b \sim
0.5$).  In \S \ref{subsec:T0-gamma-constraints} we will use the spike
width distribution to constrain the temperature of the IGM.

\subsubsection{The correlation of spike  width and height:  $\log b$ vs \logaNHI}
\label{subsec:b-NHI-corr-cold-hot}

\InputFigCombine{log_b_vs_log_NHI_L40N2048_COLD_HOT_ATON_PATCHY.pdf}{180}{%
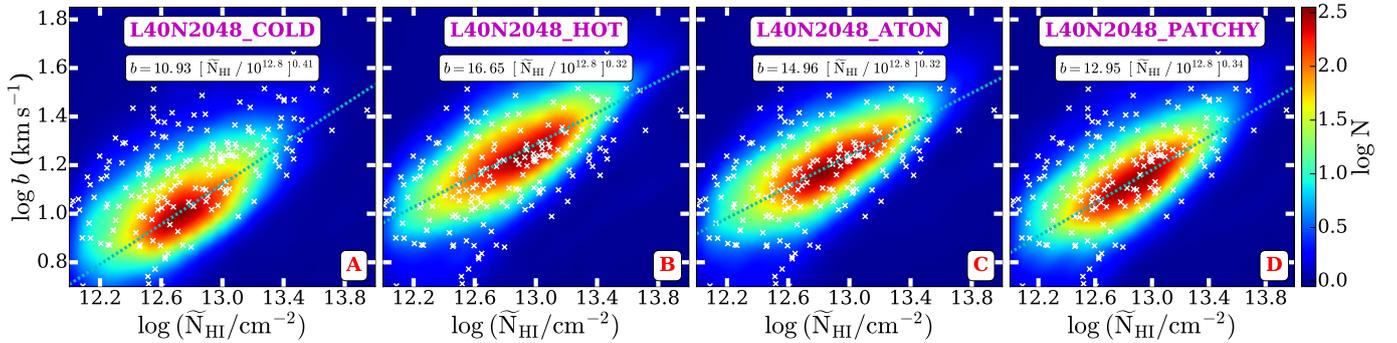
  Panels A, B, C and D show the correlation of the line-width
  parameter ($\log \: b$) with the pseudo-column density (\logaNHI) of
  the transmission spikes in, respectively, the \cold, \hot, \aton and
  \patchy models at $5.5 < z < 5.7$.  The correlation is fitted with a
  straight line (cyan dotted line).  The $\log \: b$ parameter at
  fixed \logaNHI is systematically larger in the hot model ($b \sim
  16.65$ \kmps at \logaNHI$=12.8$) compared to the \cold model ($b
  \sim 10.93$ \kmps at \logaNHI$=12.8$).  The slope of the correlation
  is steeper for the \cold ($\sim 0.41$) model compared to the \hot
  model ($\sim 0.32$).  The $b$ parameters in the \aton model ($b \sim
  14.96$ \kmps at \logaNHI$=12.8$) are systematically larger than in
  the \patchy model ($b \sim 12.95$ \kmps at \logaNHI$=12.8$).  This
  is because temperatures in the \aton model are larger than in the
  \patchy models for $\Delta < 1$ (see
  \figref{fig:T-b-corr-cold-hot-aton-patchy}).  The white crosses show
  the scatter in $\log \: b$ and \logaNHI in the observed spectra.  We
  discuss the ($\log \: b$) vs \logaNHI correlation for optically thin
  and radiative transfer simulations in \S
  \ref{subsec:b-NHI-corr-cold-hot} and \S
  \ref{subsec:b-NHI-corr-aton-patchy} respectively.
}{\label{fig:b-NHI-corr-cold-hot-aton-patchy}}

The relation of absorption line widths with column density and its
relation with the thermal state of the gas at $z<4$ has been widely
discussed in the literature
\citep{schaye1999,bolton2014,gaikwad2017b,rorai2017b,rorai2018,hiss2018,hiss2019}.
The equivalent relation for the Voigt profile parameters of the
transmission spikes in our simulated spectra is compared in
\figref{fig:b-NHI-corr-cold-hot-aton-patchy} to the observed spectra
(white crosses). \figref{fig:b-NHI-corr-cold-hot-aton-patchy} shows a
strong positive correlation between $\log b$ and \logaNHI for both
models.  We fit this correlation with a straight line of the form $b =
b_0 \; [{\rm \widetilde{N}_{HI}} \; / \;10^{12.8}]^{\alpha}$ with $b_0
= (16.65, 10.93) $ \kmps) and $\alpha = (0.32, 0.41)$) for the \hot
and \cold model, respectively.  The \hot model is in significantly
better agreement with the observations than the \cold model.  The
somewhat flatter slope in the \hot model is likely due the flatter TDR
in the \hot simulation.  Note further that the scatter in \logaNHI is
slightly larger in the \hot model.  This is because the spikes are
more blended and hence less distinctive
(\figref{fig:viper-fit-example} and \figref{fig:los-comparison}).  The
scatter in $\log b$ is similar.

In summary, we find strong correlations of the Voigt profile
parameters with physical quantities in the optically thin simulations,
where : (i) the gas probed by the transmission spikes is typically
underdense ($\Delta \sim 0.3)$ and (ii) the spike widths (heights) are
strongly (anti-) correlated with temperature (density).

\section{Comparing transmission spike properties in observations and simulations}
\label{sec:spike-statistics}

\subsection{Characterising the flux distribution in transmission spikes}

Constraints on cosmological and astrophysical parameters from \lya
forest data have been obtained typically by using either a variety of
statistical measures of the \lya transmitted flux and/or Voigt profile
decomposition
\citep{storrie1996,penton2000,mcdonald2000,mcdonald2005,viel2004b,viel2009,becker2011,shull2012b}.
Here we briefly consider both approaches before focusing on
constraints on the thermal state of the IGM from the width
distribution of transmission spikes.  The transmitted flux based
statistics (such as the probability distribution function and power
spectrum) are straightforward to derive from simulations and
observations.  For the Voigt profile parameter based statistics, we
have fitted Voigt profiles to the inverted transmitted flux, $1-F$,
using \viper \footnote{Note that  as the transmission  spikes are generally not 
``saturated'' in 1-F, we are effectively  fitting Gaussian profiles.}. 

The simulated spectra mimic the observed spectra in
terms of \SNR and instrumental resolution.  \viper accounts for these
effects when determining the best fit parameters, the $1 \sigma$
statistical uncertainty on best fit parameters and a significance
level for each Voigt component. For deriving the spike statistics, we
chose only those Voigt components with relative error on parameters
$\leq 0.5$ and with a significance level $\geq 3$.  We consider three
statistics of the flux distribution in the transmission spikes that
are sensitive to astrophysical parameters i.e., \GHI, $T_0$ and
$\gamma$ (for a given cosmology) which are discussed below.

\subsection{Statistics of the flux distribution in transmission spikes}

\subsubsection{Spike width ($b$-parameter) distribution function} 

For absorption features the line width distribution is frequently used
as a diagnostic for the gas temperature, turbulence, and the impact of
stellar and AGN feedback on the IGM at $z<5$
\citep{rauch1996,tripp2008,oppenheimer2009,muzahid2012,viel2016,gaikwad2017a,nasir2017}.
By contrast, the width of the individual spikes is not a direct
measure of the temperature of the low density gas (i.e., $b_{\rm
  spike} \neq \sqrt{2 k_{\rm B} T/m} $).  Nevertheless we find that the widths
of transmission spike components is systematically larger if the
temperature of the IGM is larger. \footnote{Since spikes trace cosmic
  voids, the spike width distribution could in principle also be
  sensitive to cosmological parameters e.g., $h$, $\Omega_{\Lambda}$,
  $\sigma_8$ and $n_s$. In this work we used the spike width
  distribution to constrain the thermal state of IGM for a given
  cosmology.}

\subsubsection{Pseudo Column Density Distribution Function (pCDDF)} 

Similar to the \HI column density distribution function (CDDF) at low
redshift, we define the pseudo-CDDF (pCDDF) as the number of spikes
with a pseudo column density in the range \logaNHI to $\delta$\logaNHI
in the redshift interval $z$ to $z + \delta z$
\citep{schaye2000,shull2012b}.  We calculate the pCDDF in 7 \logaNHI
bins centered at $12.7, 13.1,\cdots,13.9$ with \dlogaNHI$=0.2$. This
choice of bins is motivated by the \SNR and resolution of the observed
spectra. The pCDDF characterizes the height and number of spikes in a
given redshift bin, and is sensitive to the thermal and ionization
parameters of the IGM.

\subsubsection{Transmitted flux power spectrum (FPS)} 

The transmitted flux power spectrum (FPS) is frequently used to
constrain cosmological \citep{mcdonald2000,meiksin2004,viel2004b} and
astrophysical parameters \citep[especially parameters describing the
  thermal state;][]{walther2019,boera2019}.  The FPS is a measure of
the clustering of the pixels in transmission spikes.  One can also
study the two point correlation of spikes \citep[see e.g.,
][]{maitra2019}.  However, due to the limited number of observed QSOs
and the smaller number of spikes detected per sightline, the two point
correlation function of spikes is rather noisy.  It is important to
note that, unlike the pCDDF or the spike width distribution, the FPS
is a transmitted flux based statistic that does not require us to fit
the spikes. The FPS can, however, only be reliably estimated for a
limited range of scales because of finite length of the spectra and
other systematic effects. The smallest $k$ (larger scales) modes are
limited by the length of the simulation box and continuum fitting
uncertainties of the observed spectra. The largest $k$ modes (smallest
scales) on the other hand are limited by the resolution of the
instrument and the noise properties (\SNR and noise correlation scale
if non-Gaussian) of the observed spectra. To account for this, we
calculate the FPS in the range $0.01 \leq k \: ({\rm s \: km^{-1}})
\leq 0.237$ with bin width $\delta \log k = 0.125$ \citep{kim2004}\footnote{
The smallest $k$ mode corresponds to a scale of $\sim 10 h^{-1}$ cMpc.}.

\subsubsection{Error estimation} 
We estimate the error for the spike and flux statistics from the simulation using
an approach similar to that in \citet{rollinde2005} and \citet{gaikwad2018}.  For this,
we generate samples of 5 simulated \lya forest spectra corresponding to 5
observed QSO spectra     with the same observational property i.e., redshift
path length, noise property, number of pixels etc. A collection of 5
spectra constitutes a single mock sample.  We generate 80 such mock
sample and compute the covariance matrix for each statistics. We find that
the covariance matrix is converged and dominated by diagonal terms for all the
statistics.  
    
We have also estimated the covariance matrix from observations using a
bootstrap method.  Here we find that the off-diagonal terms of the
covariance matrix estimated with the bootstrap method are not converged.  The
bootstrap method (diagonal terms) underestimates the error by $\sim 15$
percent as compared to those estimated from the simulations. Throughout this
work, we use bootstrap errors increased by a corresponding factor for each
statistics.

\InputFigCombine{{FPS_CDDF_BPDF_All_z_COLD_HOT_DEFAULT}.pdf}{165}{
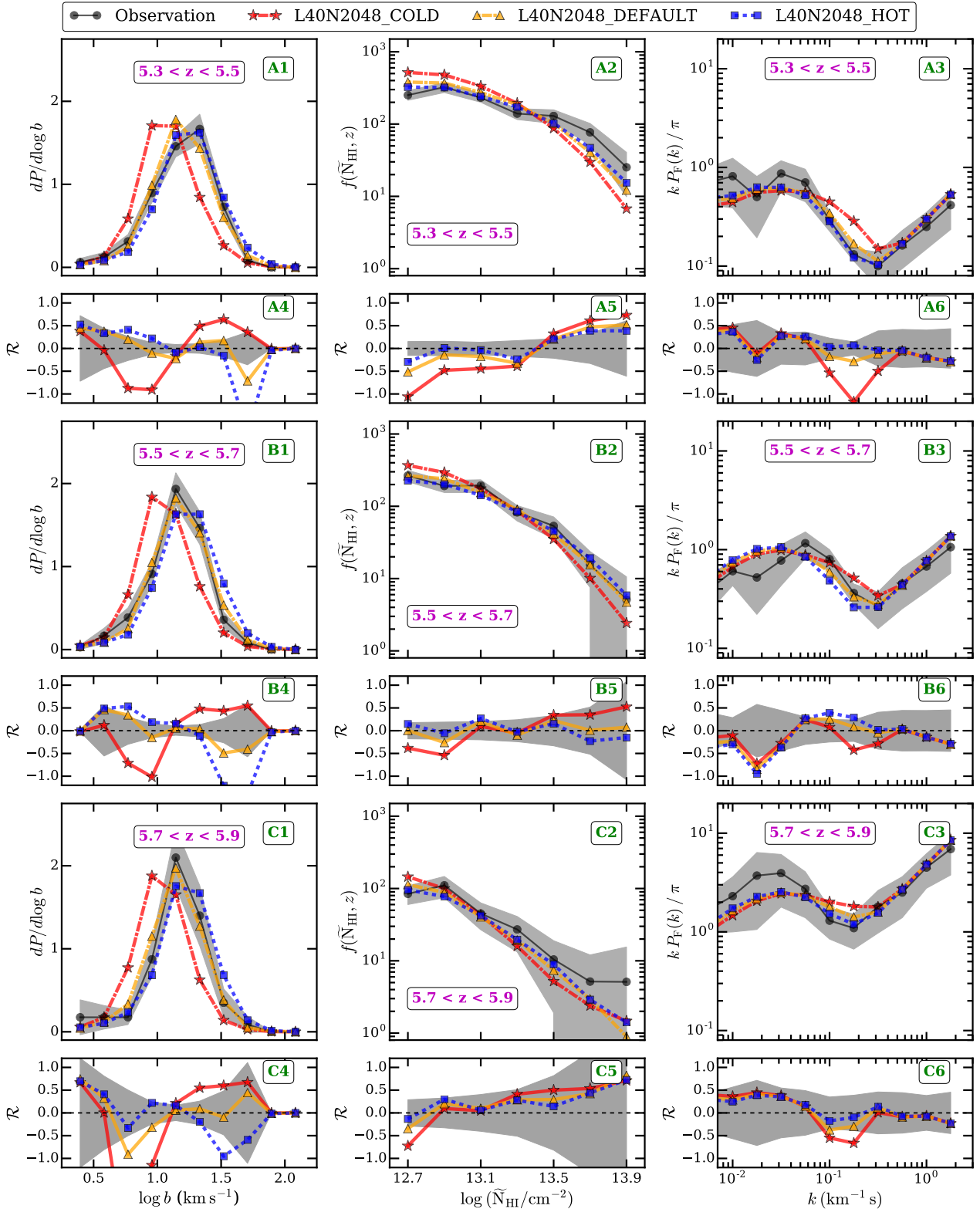
  Panels A1, A2 and A3 show a comparison of spike width distribution,
  pCDDF and FPS from observations (black circles) with those from
  \cold (red stars), \default (orange triangles) and \hot (blue
  squares) optically thin simulations at $5.3 < z < 5.5$. The
  $1\sigma$ uncertainties are estimated from the simulated spectra and
  are shown by the grey shaded regions.  Panels A4, A5 and A6 show the
  corresponding residuals between the models and observations. The
  gray shaded region in panels A4-A6 corresponds to the gray shaded region 
  in panels A1-A3. Panels  B1-B6 and C1-C6 are similar to panels A1-A6 
  except for the different
  redshift range, $5.5 < z <5.7$ and $5.7 < z < 5.9$, respectively.
  Note that the \hot and \default models are in better agreement with
  observations than the \cold model at all redshifts. The agreement between 
  model and observed data is discussed quantitatively in appendix 
  \ref{app:statistics-comparison}. 
  The spike statistics in the \default model are very close to those of the best
  fit model obtained by varying $T_0$ and $\gamma$ in \S
  \ref{subsec:T0-gamma-constraints}.
}{\label{fig:fps-cddf-bpdf-cold-hot}}

\subsection{Transmission spike statistics: optically thin simulations vs observations}
\label{sec:result-OT}

\figref{fig:fps-cddf-bpdf-cold-hot} compares the spike width
distribution, pCDDF and FPS from observations with that of the
optically thin simulations  for three different
redshift bins centered at $z=(5.4,5.6,5.8)$.\footnote{The mean
  transmitted flux in \hot, \cold and \default models is matched to that in the observed
  spectra to account for the uncertainty in the continuum placement
  and UV background amplitude, see appendix
  \ref{app:observational-systematics}.}  The corresponding residuals suggest a good level of agreement between the \default
model and observations for all three statistics.  The residuals for the  \hot model with $T_0 \sim 11000$ K and $\gamma \sim 1.2$  
are somewhat larger.  Overall the three statistics for  the \default and \hot   model 
are in  noticeably better agreement than those for the   \cold model, with differences in the Doppler parameter distribution being most pronounced where agreement with the \default model is significant better than with 
both the \hot and \cold model.  The \cold model  also predicts more power on small scales, as well as a larger number of
spikes with small \logaNHI than observed,  and is clearly  the model  that is least  consistent with the observations.  
We further quantify the degree of agreement between models and observation in appendix \ref{app:statistics-comparison}.

\subsection{Constraining thermal parameters with optically thin simulations} 
\label{subsec:T0-gamma-constraints}

In the optically thin simulations there is a well defined TDR in
underdense and moderately overdense regions. It is thus common
practice to study the effect of the thermal state on flux statistics
by imposing different TDRs by rescaling temperatures
\citep{hui1997,rorai2017b,rorai2018}.  
In Appendix~\ref{app:spike-statistics-sensitivity} we use such
rescaling to show how the three flux statistics we consider here
depend on thermal parameters $T_0$ and $\gamma$ and the
photo-ionization rate, and demonstrate that it is the width
distribution of the transmission spikes which is most sensitive to the
thermal state of the gas.  We further show that, for the reionization
and thermal history models we consider in this work, the spike
statistics are only very weakly affected by pressure (Jeans) smoothing
at the typical gas densities probed by the \lya transmission spikes
\citep{gnedin1998,theuns2000a,peeples2010,girish2015,lukic2015,nasir2016,
maitra2019,wu2019}. 

We use the spike width distribution to obtain an estimate of the best
fit values and uncertainty for the thermal parameters \footnote{We
  find that the FPS and pCDDF statistics are sensitive to continuum
  fitting uncertainty and \GHI, whereas the spike width distribution
  is less sensitive to continuum placement and \GHI (see Appendix
  \ref{app:observational-systematics})}.  We vary $T_0$ and $\gamma$
by assuming a TDR of the form $T = T_0 \: \Delta^{\gamma-1}$ for
$\Delta < 10$ and $T=T_0 \: 10^{\gamma-1}$ for $\Delta \geq 10$.  We
use the $\Delta$ and $v$ fields from the optically thin \default
model, which falls in between the \cold and \hot models. We vary $T_0$
between $6000\rm\, K$ to $20000\rm\, K$ in steps of $500 \: {\rm K}$
and $\gamma$ between $0.4$ to $2.0$ in steps of $0.05$. For each model
we (i) compute the \lya transmitted flux, (ii) post-process the
transmitted flux to match observations, (iii) fit the inverted
transmitted flux with Voigt profiles and (iv) compute spike
statistics. We use 2000 sightlines for each model.  
\figref{fig:T0-gamma-constraints} shows the $1\sigma$ constraints on
$T_0-\gamma$ in three redshift bins.\footnote{To a good approximation, the
    likelihood function is Gaussian distributed on $T_0-\gamma$ grids.  See
    Appendix \ref{app:resolution-study} for details.}
\figref{fig:T0-gamma-constraints} also shows the marginal distributions of $T_0$
and $\gamma$.  \tabref{tab:T0-gamma-constraints} summarizes the best fit values
and uncertainty on $T_0$ and $\gamma$ values under the assumption that the IGM
is optically thin. The uncertainty on the $T_0$ and $\gamma$ measurement accounts
for the uncertainty due to continuum fitting, mean flux and Jeans smoothing
effects (see Appendix \ref{app:observational-systematics} and
\ref{app:spike-statistics-sensitivity}).  The best fit values and uncertainty
on $T_0$ and $\gamma$ are consistent with each other within $1\sigma$ for the three redshift bins.
As the spikes are probing predominantly gas at densities lower than the
mean, the inferred values for  $T_0$ and $\gamma$  are strongly  correlated. We
will come back to this in Appendix~\ref{app:spike-statistics-sensitivity}.

\figref{fig:T0-gamma-evolution} compares the evolution of $T_0$ and
$\gamma$ from this work with a variety of measurements in the
literature
\citep{becker2011,bolton2012,bolton2014,boera2014,rorai2017b,hiss2018,walther2019,boera2019}.
\footnote{\citet{bolton2012} measure $T_0$ in QSO proximity regions at
  $z\sim 6$. Their $T_0$ constraint including (excluding) a model
  for the \HeII photo-heating by the QSOs is shown by green circles
  (blue squares) in \figref{fig:T0-gamma-evolution}.}  Theoretical
  models for the evolution of $T_0$ and $\gamma$ from
  \citet{puchwein2019} and \citet{haardt2012} are shown by
  the dashed and dotted curves, respectively\footnote{The $T_0$ and 
  $\gamma$ evolution in \citet{khaire2015a,khaire2019a} and \citet{faucher2019} 
  are similar to that in the \citet{haardt2012} UVB model at $z > 5$.}.  The $T_0$ and $\gamma$
  evolution in these models is obtained for a uniform but time
  evolving UVB and assuming non-equilibrium ionization evolution. Our
  $T_0$ and $\gamma$ constraints in the redshift range $5.3<z<5.9$ are
  consistent with the corresponding evolution of the \default model in
  \citet{puchwein2019} within $1
  \sigma$. 

\figref{fig:T0-gamma-evolution} also shows that the $T_0$ evolution in
the \hot (\cold) model is systematically larger (smaller) than in the
corresponding \default model.  The errors on $T_0$ and $\gamma$
account for the statistical and systematic uncertainty (mainly due to
continuum fitting).  It is interesting to note that the uncertainty
on our $T_0$ measurement is smaller than the $T_0$ evolution
spanned by the \hot and \cold models.  Our $T_0$ ($\gamma$)
measurements are higher (lower) than the measurement of
\citet{walther2019}. Note, however, that the $T_0$ and $\gamma$
constraints in \citet{walther2019} are obtained using the FPS whereas
we obtained the $T_0$ and $\gamma$ constraints from the spike width
distribution that is less sensitive to continuum placement and \GHI
uncertainty (see Appendix
\ref{app:observational-systematics}). Further note that the best fit
$T_0$ and $\gamma$ values obtained for the three redshift bins are
close to those in the \default optically thin
simulation. 

\figref{fig:fps-cddf-bpdf-cold-hot} also shows that the
FPS and pCDDF statistics of the \default model are consistent with the
observations within 1.5 $\sigma$.  The best fit model obtained by
matching the spike width distribution with observations should
therefore also have FPS and pCDDF statistics in good agreement with
observations. Our $\gamma$ constraints ($\gamma \sim 1.2$) correspond
to a TDR that is moderately steeper than isothermal.  However, as we
show later, there is no single power law TDR at the redshift of our
analysis since the reionization is a patchy inhomogeneous process with
different regions reionising at different times. At $5.3 < z < 5.9$,
$\gamma$ is therefore not well defined in our RT simulations
\citep{keating2018}.  In summary, the $T_0$ ($\gamma$) constraints
obtained in this work are larger (smaller) than those obtained by
\citep{walther2019}.  We do not see a significant evolution of $T_0$
and $\gamma$ in the redshift range $5.3 < z < 5.9$.

\begin{table}
\centering
\caption{Constraints on $T_0-\gamma$ from optically thin simulations}
\begin{tabular}{ccc}
\hline \hline
Redshift & $T_0 \pm \delta T_0$ & $\gamma \pm \delta \gamma$ \\ 
\hline
$5.3 < z < 5.5$ & $11000 \pm 1600$ & $1.20 \pm 0.18$ \\
$5.5 < z < 5.7$ & $10500 \pm 2100$ & $1.28 \pm 0.19$ \\
$5.7 < z < 5.9$ & $12000 \pm 2200$ & $1.04 \pm 0.22$ \\
\hline \hline
\end{tabular}
\label{tab:T0-gamma-constraints}
\end{table}

Transmission spikes in optically thin simulations are mainly produced
by fluctuations in the density field and the effect of peculiar
velocities.  Furthermore, the temperature is strongly and tightly
correlated with density in optically thin simulations.  However, at
the redshifts considered here this almost is certainly not realistic.
One would expect a large scatter in temperature for a given density as
the reionization process will be inhomogeneous, with different regions
ionized at different times \citep{abel1999,miralda2000,trac2008,
  choudhury2009}.  The resulting spatial fluctuations in the amplitude
of the UVB and the TDR are not present in optically thin
simulations\footnote{For optically thin simulations, the variation in
  temperature along a sightline is closely coupled to the  variation in the density
  field.}.  However, as shown in \S
\ref{sec:spike-theoretical-analysis}, transmission spikes can also be
produced by fluctuations in the UVB amplitude and/or
temperature. Including these radiative transfer (RT) effects is
therefore particularly relevant if reionization ends as late as
suggested by the large spatial fluctuations in the \lya forest opacity
\citep{keating2019,kulkarni2019a}.

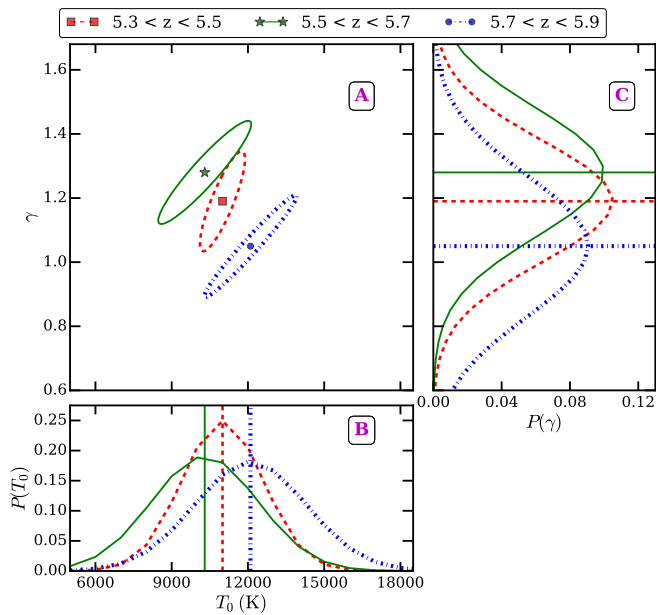
\InputFig{T0_gamma_constraints_KDE.pdf}{85}{Panel A shows $1\sigma$
  constraints on $T_0$ and $\gamma$ obtained by comparing the
  transmission spike width distribution from optically thin
  simulations with observations at $5.3 < z < 5.5$ (red dashed curve),
  $5.5 < z < 5.7$ (green solid curve) and $5.7 < z < 5.9$ (blue dotted
  curve). $T_0$ and $\gamma$ are varied in post-processing assuming a
  power-law TDR (the effect of Jeans smoothing is small, see
  \figref{fig:param-effect-L-jeans} in the appendix).  Panel B and C
  shows the marginal distributions for $T_0$ and $\gamma$
  respectively.  The best fit $T_0$ and $\gamma$ for $5.3 < z < 5.5$
  are shown by the red square in panel A and red dashed lines in panel
  B and C, respectively.  Corresponding best fit values for $5.5 < z <
  5.7$ and $5.7 < z < 5.9$ are shown by the green star and solid line
  and blue circle and dotted line, respectively.
}{\label{fig:T0-gamma-constraints}}

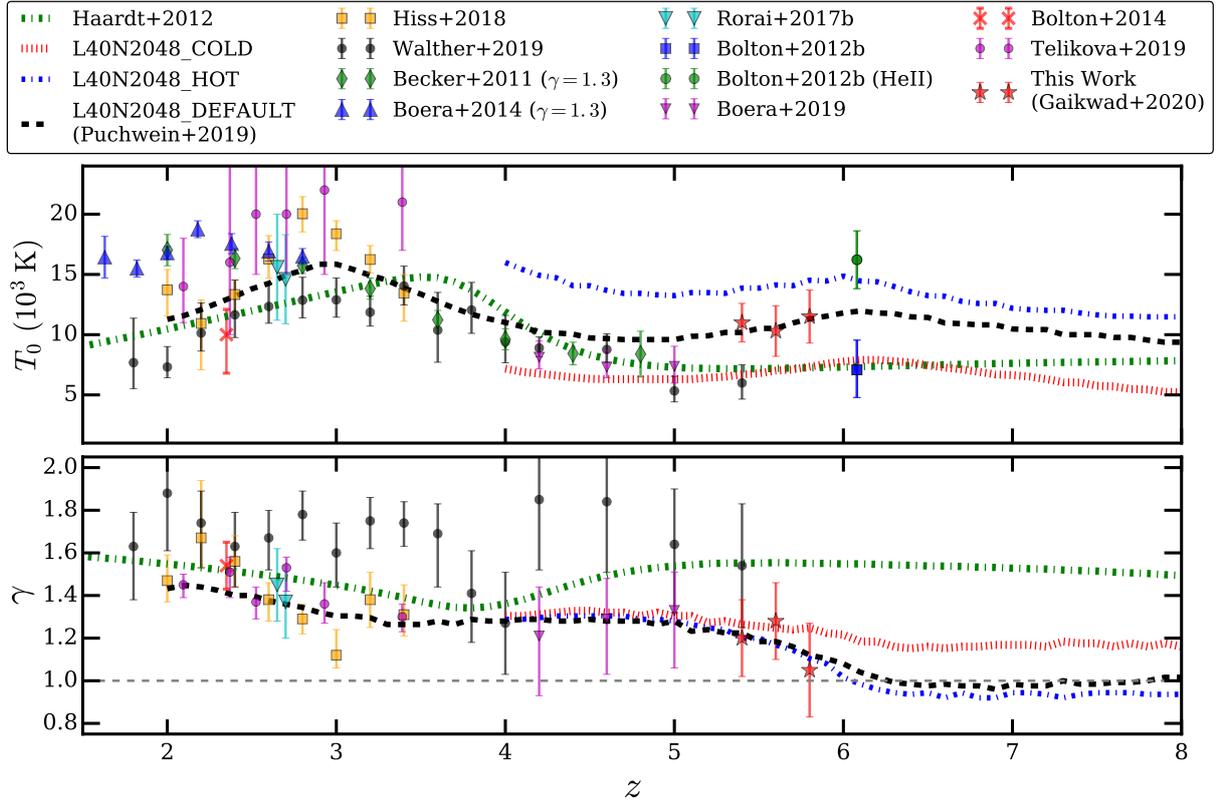
\InputFigCombine{Thermal_Parameter_Evolution.pdf}{160}{The evolution
  of thermal parameters $T_0$ and $\gamma$ from the literature and in
  this work (red stars with error bars) are shown in the top and
  bottom panels, respectively
  \citep{becker2011,bolton2012,bolton2014,boera2014,rorai2017b,hiss2018,walther2019,boera2019,telikova2019}.
  Note that the temperature constraints from \citet{becker2011} and
  \citet{boera2014} are not measured at the mean density, and have
  therefore been scaled to a $T_{0}$ value assuming a TDR with slope
  $\gamma=1.3$.  The $T_0$ and $\gamma$ evolution in the \hot,
  \default and \cold \xspace \relics simulations is shown by the blue
  dash-dotted, black dashed and red dotted curves respectively.  The
  corresponding $T_0$ and $\gamma$ evolution in the \citet{haardt2012}
  UVB synthesis model is shown by a green dotted curve.  This $T_0$ and
  $\gamma$ evolution is obtained by assuming a uniform UVB and solving
  for the non-equilibrium ionization evolution
  \citep{haardt2012,puchwein2019}.  Our constraints on $T_0$ and
  $\gamma$ are consistent within $1 \sigma$ with \citet{puchwein2019}.
}{\label{fig:T0-gamma-evolution}}

\section{Full radiative transfer simulations}
\label{sec:spikes-in-RT-simulation}
\subsection{The radiative transfer simulations in the \relics simulation suite}

In addition to the optically thin simulations discussed in
\S\ref{subsec:sherwood_relics}, we have also performed post-processed
radiative transfer simulations and hybrid radiative
transfer/hydrodynamical simulations as part of the \relics simulation
suite. The former simulations model patchy reionization by performing
the radiative transfer in post processing on optically thin
simulations. This captures many aspects of patchy reionization such as
large spatial fluctuations in the photo-ionization rate and
temperature, but misses the hydrodynamic response of the IGM to the
heating and can thus not accurately predict spatial variations in the
pressure smoothing or the distribution of shock heated gas. The hybrid
simulations aim to capture these aspects as well.

The post-processed radiative transfer simulations were performed with
the GPU-accelerated \atoncode code \citet{aubert2008}, which uses a
moment based radiative transfer scheme along with the M1 closure
relation. The advection of the radiation was performed using the full
speed of light and a single frequency bin for all ionizing
photons. Ionizing sources were inserted into dark matter halos as in
\citet{kulkarni2019a}. The (mean) energy of ionizing photons was
assumed to be $18.6$ eV. In the following, we will refer to the
post-processed radiative transfer simulation performed on top of our
\default optically thin simulation by the term \aton simulation.  It
used $2048^3$ cells in the $(40 \, h^{-1} \, \textrm{Mpc})^3$ box and
hydrogen reionization completes at $z \approx 5.2$, consistent with
the late reionization history found to be favoured by large scale \lya
forest fluctuations \citep{kulkarni2019a}.

The hybrid radiative transfer/hydrodynamical simulation, referred to
as the \patchy simulation, takes the reionization redshift and \HI
photo-ionization rate maps produced in the \aton simulation as inputs.
These are fed to our modified version of \pgthree, where they are used
in the non-equilibrium thermochemistry solver instead of an external
homogeneous UV background model.  To obtain consistent density and
radiation fields, we use the same initial conditions as in the
\default optically thin simulation on which the \aton run is based. At
each timestep, for each SPH particle we check whether it resides in a
region in which reionization has already begun. This is assumed to be
the case if the ionized fraction in the corresponding cell of the
\aton simulation has exceeded 3 percent. All particles located in such
regions are assumed to be exposed to an ionizing radiation field,
which is obtained by interpolating the \GHI maps produced by \aton in
redshift and reading out the value of the cell containing the
particle.  This value is then adopted for \GHI in the non-equilibrium
thermochemistry solver. The \HI photo-heating rate is computed from
\GHI assuming the same mean ionizing photon energy, $18.6$ eV, as in
\aton. As we do not follow \HeI and \HeII ionizing radiation
separately, we use a few simple assumptions to set their
photo-ionization and heating rates. For \HeI we use the same
photo-ionization rate as for \HI, but we adopt a photo-heating rate that
is 30 percent larger than that of \HI. For \HeII, we use the rates of
the \textit{fiducial} UV background model of
\citet{puchwein2019}. This hybrid method results in ionized regions
and inhomogeneous photo-heating that closely match those in the parent
radiative transfer run, while at the same time following the
hydrodynamics and hence including consistent pressure smoothing, as
well as shock heating.

\subsection{The thermal state of the gas in full radiative transfer
simulations} \label{subsec:tdr-aton-patchy}

The main difference in the radiative transfer simulation is that there
are large spatial variations in the TDR
\citep{trac2008,keating2018,kulkarni2019a}.  In panels C and D of
\figref{fig:tdr-cold-hot-aton-patchy} we can compare the TDRs from the
\aton and \patchy RT simulations to the optically thin simulations in
panels A and B.  Unlike the optically thin simulations, the TDR in the
radiative transfer simulations at $\Delta < 10$ can not be described
by a single power law \citep{bolton2004}.  
The regions that have been ionized most recently have a flat
TDR while regions that have been ionized earlier have progressively
steeper TDRs.  In the redshift range considered here there are also
still significant spatial fluctuations in the amplitude of the UVB. As
a result, the variations in temperature for a given $\Delta < 10$ is
large.  A crucial difference between the \aton and \patchy simulations
is also evident in \figref{fig:tdr-cold-hot-aton-patchy}.  The \aton
simulation does not account self-consistently for shock heating of the
gas.  There is thus much less gas with $T>30000$ K in the \aton
simulation than in the \patchy simulation.  As a consequence there is
less gas with $\Delta < 10$ in the temperature range described by two
straight lines (see \figref{fig:tdr-cold-hot-aton-patchy}) in the
\patchy simulation.  As we will show later this has the effect of
producing slightly larger spike widths in the \aton simulations.

Table \ref{tab:simulation-details} also shows the physical effects
responsible for the occurrence of spikes in the \aton and \patchy
radiative transfer simulations.  Similar to the optically thin
simulations, most of the spikes ($\sim 89$ percent) in the \aton and
\patchy simulations occur in underdense regions and for $\sim 18$
percent of spikes the gas shows a diverging velocity field along the
sightline. However, in contrast to optically thin simulations, around
50 percent of spikes show an enhancement of \GHI and temperature.
Note here that the recent observations of high redshift Lyman alpha
emitters and Lyman break galaxies suggests that the transmission spikes are 
spatially correlated with the ionizing radiation escaping these galaxies \citep{meyer2020}.
Thus, for the transmission spikes in the full radiative transfer
simulations, all the physical processes discussed in
\S\ref{sec:spike-theoretical-analysis} and \figref{fig:spike-origin}
contribute to the occurrence of spikes.

\subsubsection{Dependence of spike height on density:  \qtau{\Delta} vs \logaNHI }
\label{subsec:delta-NHI-corr-aton-patchy}

In \figref{fig:delta-NHI-corr-cold-hot-aton-patchy} we compare the
dependence of \qtau{\Delta} on \logaNHI for the \aton and
\patchy simulations to the optically thin \cold 
and \hot simulations. Similar to the optically thin
simulations (\figref{fig:delta-NHI-corr-cold-hot-aton-patchy}),
\qtau{\Delta} and \logaNHI are anti-correlated.  The normalization
($\Delta_0$) and slope ($\beta$) are similar in both simulations.  The
$\Delta_0$ ($\beta$) in the RT simulations are slightly larger
(smaller) compared to the optically thin simulations.  The spikes
(irrespective of \logaNHI) in the RT simulations are produced from
underdensities somewhat larger than in the optically thin simulations.
This is expected, as spikes (at a given \logaNHI) in the RT
simulations are produced by all four physical effects we discussed
previously, i.e., fluctuations in density, peculiar velocity, UVB and
temperature.  \figref{fig:delta-NHI-corr-cold-hot-aton-patchy} also
shows that the scatter in \qtau{\Delta} (at a given \logaNHI) is
larger in the radiative transfer simulations due to fluctuations in
UVB, temperature and pressure smoothing effects.  Note, however, that
in all simulations the spikes occur in underdense regions with $\Delta
< 1$.

\subsubsection{Dependence of temperature on spike width: \qtau{T} vs $b$}
\label{subsec:T-b-corr-aton-patchy}

\figref{fig:T-b-corr-cold-hot-aton-patchy} compares the dependence of
optical depth weighted temperature (\qtau{T}) on spike widths
($b$-parameter) for the RT simulations with that in
the optically thin simulations.  Unlike the optically
thin simulations, the RT simulations show a large scatter in
temperature for a given spike width due to fluctuations in the UVB
amplitude and temperature.  Furthermore, the temperature in the
\patchy simulation is smaller than in the corresponding \aton
simulation.  This is because (i) the amount of gas with $\Delta < 1$
and $T < 30000 \: {\rm K}$ is larger in the \patchy simulation and (ii) due
to the post-processed nature of the \aton simulation the (adiabatic)
change in the temperature due to changes in density (the $d\Delta /dt$
term) is not accounted. As a result, the spike widths are also
slightly smaller in the patchy simulations.
\subsubsection{The relation of spike width and height: $b$ vs \logaNHI}
\label{subsec:b-NHI-corr-aton-patchy}

\figref{fig:b-NHI-corr-cold-hot-aton-patchy} compares the $b-$\logaNHI
correlation for the \aton and \patchy simulations
to that in the optically thin simulations. Similar to
the optically thin simulations, $\log b$ and \logaNHI are strongly
anti-correlated in the RT simulations.  The normalization of the
correlation $b_0$ is smaller in the \patchy ($\sim 12.95$ \kmps)
simulation than in the \aton ($\sim 14.96$ \kmps) simulation due to
the smaller temperature of the gas probed by the transmission spikes
in the latter.  The slope of the correlation ($\alpha = 0.32$ for
\aton and $\alpha =0.34$ for \patchy) and the scatter in the TDR (at
$\Delta < 1$ in \figref{fig:tdr-cold-hot-aton-patchy}) is relatively
similar for both simulations.  It is interesting to note here that
$\alpha$ in the RT simulations is similar to that in the \hot
optically thin simulation, while $b_0$ in the RT simulations is
smaller than in the \hot optically thin simulations.  The smaller
value of $b_0$ in the RT simulation is a consequence of fluctuations
in UVB amplitude and temperature.

In summary, the RT simulations include the effects of fluctuations
in the UVB amplitude and temperature which are missing in the
optically thin simulations. Due to these effects: (i) the TDR in RT
simulations cannot be described by a single power-law, (ii) the
typical densities responsible for transmission spikes in the RT
simulations ($\Delta_0 \sim 0.33$) is slightly larger than in
optically thin simulations ($\Delta_0 \sim 0.25$) and (iii) the
scatter in temperature and spike width are larger.

\subsection{Transmission spike properties: Full radiative transfer simulations vs observations}
\label{sec:result-RT}

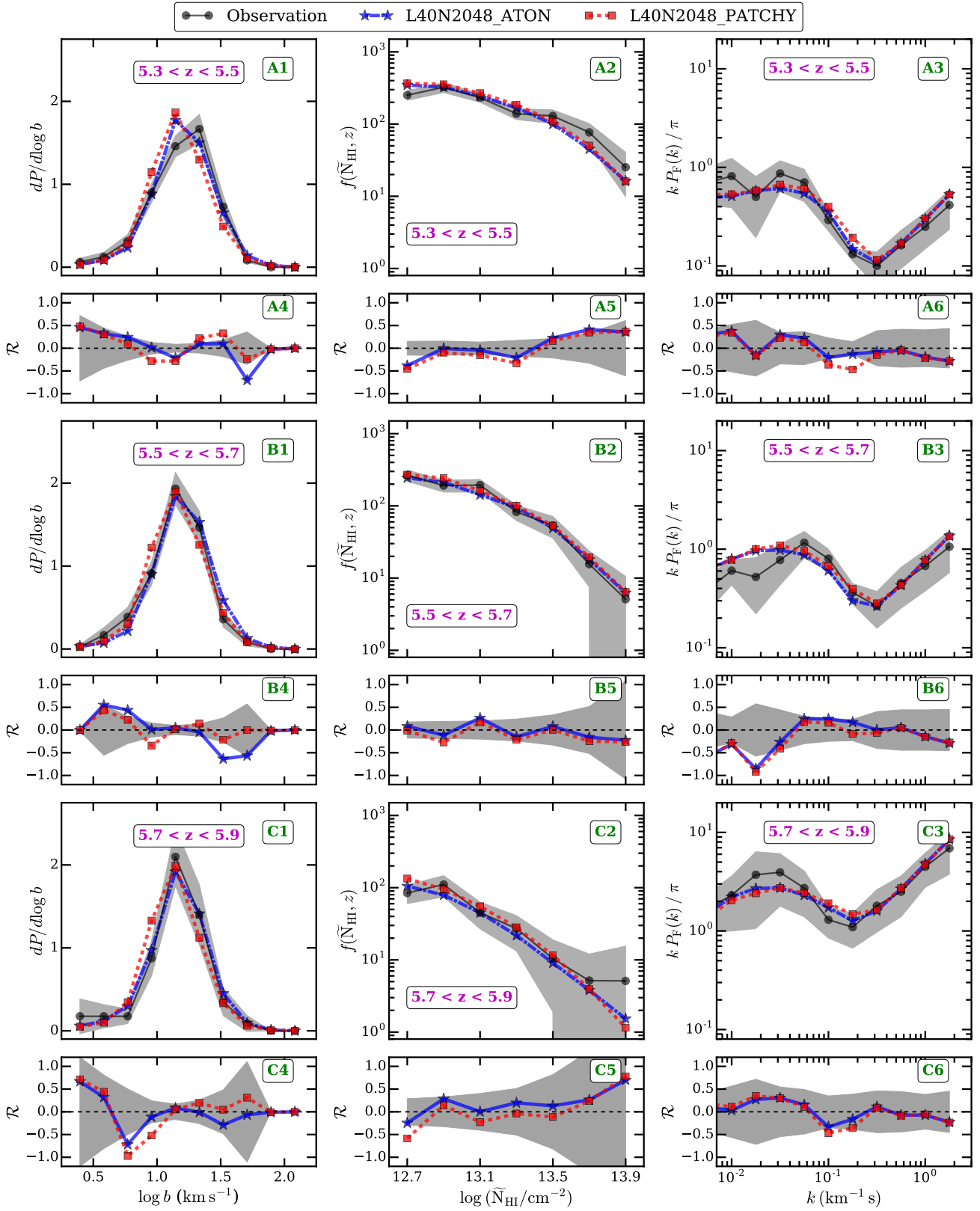
\InputFigCombine{{FPS_CDDF_BPDF_All_z_ATON_PATCHY}.pdf}{170}{Similar
  to \figref{fig:fps-cddf-bpdf-cold-hot}, except the comparison of
  spike width distribution, pCDDF and FPS is shown for the \aton (blue
  stars) and \patchy (red stars) RT simulations.  The statistics from
  the RT simulations are in better agreement with observations
  compared to those from the optically thin simulations
  (\figref{fig:fps-cddf-bpdf-cold-hot}) at all redshifts (see \S
  \ref{sec:result-RT}).  }{\label{fig:fps-cddf-bpdf-aton-patchy}}

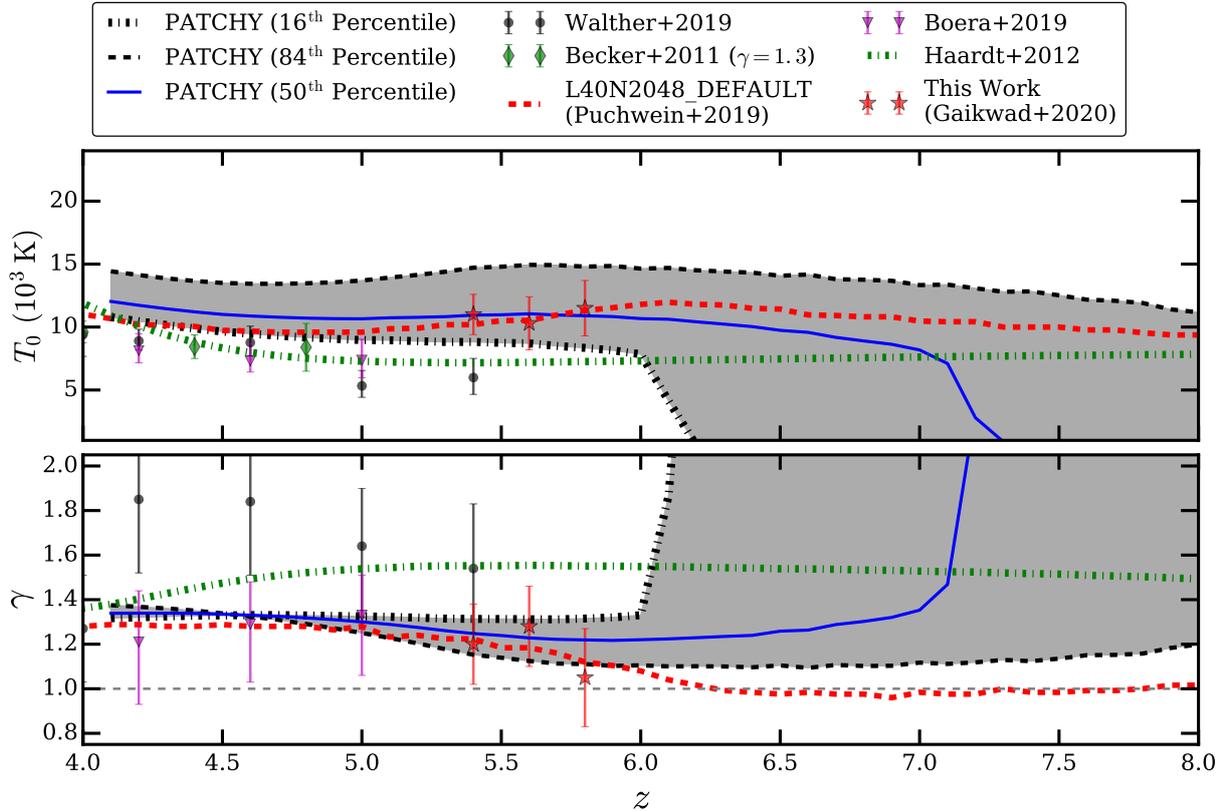
\InputFigCombine{Thermal_Parameter_Evolution_RT.pdf}{160}{Same as
  \figref{fig:T0-gamma-evolution}, except the $T_0$ and $\gamma$ evolution is
  shown from the \patchy RT simulation (shaded region) at $4 \leq z \leq 8$.
  Since there is no single power-law TDR in the RT simulations (see
  \figref{fig:tdr-cold-hot-aton-patchy}), the shaded region displays the
  16$^{\rm th}$ and 84$^{\rm th}$ percentiles of $T_0$ and $\gamma$ (see \S
  \ref{sec:result-RT}). For comparison, we also show the $T_0$ and $\gamma$
      evolution from RT models (50$^{\rm th}$ percentile, blue solid line), the
      uniform UVB \default (red dashed line) and \citet{haardt2012} (green
      dotted line) model.  The $T_0$ and $\gamma$ measured by comparing
      the observed spike width distribution with that from optically thin
  simulations are in good agreement with the median $T_0$ and $\gamma$ evolution
  from RT simulations.
  }{\label{fig:T0-gamma-evolution-RT}}

We now compare the three statistics from the RT simulations with observations
in \figref{fig:fps-cddf-bpdf-aton-patchy}. Each panel is similar to
\figref{fig:fps-cddf-bpdf-cold-hot}, except we now use the \aton and \patchy
simulations. Similar to the optically thin simulations, the mean transmitted
    flux in the \aton and \patchy models is matched to that in the observed spectra
    to account for the uncertainty in the continuum placement and UV background
    amplitude. Note that this rescaling  has surprisingly little effect on 
    the width of the transmission
    spikes (see appendix \ref{app:observational-systematics}).
Comparison of \figref{fig:fps-cddf-bpdf-aton-patchy} with
\figref{fig:fps-cddf-bpdf-cold-hot} shows that the RT simulations are in perhaps 
even better agreement with the observations than the optically thin simulations at
all redshifts. Note, however,  that  there is still  considerable  freedom  
to adjust the overall temperature in  both the RT  and optically  thin simulation.
Unfortunately, we therefore don't think that this (marginally) better agreement should be interpreted 
as (hard) evidence for the spatial variations of the TDR predicted by our RT simulations.
 The three statistics are furthermore very similar in the \aton
and \patchy simulations at all redshifts.  The spike widths in the \aton
simulations are somewhat larger than in the \patchy simulation and are in
marginally better agreement with observations than in the \patchy simulations
(see appendix \ref{app:statistics-comparison} for a comparison of the goodness of fit
for the different models). 
As explained in the previous section, this is due to the
effect of slightly higher gas temperatures in the \aton simulations. Note,
however, also that both RT simulations are mono-frequency and the normalisation of
the temperature distribution predicted by the RT simulations is still somewhat
uncertain.

\figref{fig:T0-gamma-evolution-RT} shows the comparison of the $T_0$ and
$\gamma$ evolution with that from $T_0-\gamma$ measured assuming a
uniform UVB. Since there is no single power-law TDR in the RT
simulations (see \figref{fig:tdr-cold-hot-aton-patchy}), we show a
range in $T_0$ and $\gamma$ evolution. To obtain this range in $T_0$
and $\gamma$, we find the 16$^{\rm th}$ and 84$^{\rm th}$ percentile
temperature in four $\Delta$ bins.  We then fit a power-law TDR to the
16$^{\rm th}$ and 84$^{\rm th}$ percentile temperature values (see
Fig. \ref{fig:tdr-cold-hot-aton-patchy}\footnote{The $T_0$ and $\gamma$
for \aton and \patchy models in Fig. \ref{fig:tdr-cold-hot-aton-patchy} are 
calculated using 5$^{\rm th}$ and 95$^{\rm th}$ percentile.}). 
\figref{fig:T0-gamma-evolution-RT}
shows that the $T_0-\gamma$ constraints obtained here are consistent
with the range in the $T_0-\gamma$ evolution seen in the \patchy and
\aton RT simulations.

Thus, quite remarkably the \patchy simulation based on the self-consistent
reionization model of \citet{kulkarni2019a} that (i) simulates cosmological
density and velocity fields, (ii) includes spatial fluctuations in the UV
background and TDR, (iii) accounts for the pressure smoothing of gas and (iv)
matches the Thomson scattering optical depth \citep{planck2018}, also produces
transmission spike properties consistent with those in observed
high-resolution, high-$z$ QSO absorption spectra. Note, however, that there is
still some uncertainty in the post-reionization temperatures in RT simulations
\citep[see][Puchwein et.al. 2020, in prep]{daloisio2018}. With 18.6 eV the energy of
ionizing photos (and thus the post-reionization temperatures) of the
simulations used here fall between those  in  \citet{keating2019} [17.6 eV] and
\citet{kulkarni2019a} [20.1 eV].

\section{Conclusions}
\label{sec:conclusion}
We have explored here for the first time the use of the transmission spikes
observed in high-$z$ QSO absorption spectra  as a tool to probe the physical
state of the IGM near the tail end of hydrogen reionization. We constrain the
thermal state of the IGM at $5.3 < z < 5.9$ by comparing the properties of \lya
transmission spikes from a sample of 5 high resolution ($v_{\rm FWHM} \sim 6$
\kmps) and high \SNR ($\sim 10$) QSO absorption spectra with that from
state-of-the-art, high resolution optically thin simulations run with \gthree
\citep{springel2005} from the \sherwood and \relics suites, as well as a
simulation post-processed with the radiative transfer code \aton
\citep{aubert2008}. The main results of this work are as follows.

\begin{itemize}
    \item In full radiative transfer simulations regions with low density,
      enhancement in the photo-ionization rate \GHI, enhancement in
      temperature and a diverging peculiar velocity field along the
      line of sight can all contribute to the occurrence of
      transmission spikes at high redshift. Most of the spikes ($\sim
      90$ per cent) in optically thin and radiative transfer
      simulations occur in regions with density $\Delta <1$.
      Optically thin simulations do not account for the effect of
      enhanced temperatures in recently ionized regions and the
      resulting spatial fluctuations in the temperature-density
      relation. Due to the assumed spatially homogeneous UV background
      amplitude they also do not account for the occurrence of
      transmission spikes due to enhancements in the photo-ionization
      rate. About 50 per cent of the transmission spikes in our RT
      simulations show the effect of an enhanced \GHI and enhanced
      temperature. In the RT simulation the transmission spikes are
      often due to either hot, recently ionized, very underdense
      regions with a diverging line of sight peculiar velocity field
      or due to somewhat less underdense and colder regions with an
      enhanced photo-ionization rate.
 
    \item The width of the components fitted to the asymmetric and 
        blended transmission spikes
      are very sensitive to the instantaneous temperature of the gas
      and are significantly broader in the optically thin \hot
      simulation than in the corresponding \cold simulation.  To
      quantify this, we have fitted multi-component Voigt profiles to
      the inverted transmitted flux $1-F$ in both simulated and
      observed spectra with our automated code \viper
      \citep{gaikwad2017b}. We derive the transmitted flux power
      spectrum, pseudo column density distribution function (pCDDF)
      and spike width ($b$-parameter) distribution functions for
      simulated and observed spectra. {We show that the spike width
      distribution is the statistic that is sensitive to the
      thermal state of the IGM.}  The dependence of the shape of the
      FPS and pCDDF on the temperature of the absorbing gas is
      somewhat weaker while their normalisation is more sensitive to
      the ionization state/neutral fraction of the IGM.


    \item We associate the observable properties of spikes with the
      physical properties of gas in simulations by studying the
      \qtau{\Delta}$-$\logaNHI, \qtau{T}$-b$ and $b-$\logaNHI
      correlations.  These correlations show that the underdensity of
      gas associated with spikes is similar i.e., \qtau{\Delta}$\sim
      0.3$ at \logaNHI$\sim$12.8 in both optically thin and radiative
      transfer simulations. The spike widths in both simulations are
      sensitive to the temperature of the gas ($b\sim 10.9$ \kmps for
      $T_0 \sim 7500$ K and $b \sim 16.6$ \kmps for $T_0 \sim 14000$
      K).  However, the temperature scatter for a given density is
      larger in the radiative transfer simulation compared to the
      optically thin simulation. As a result, a significant fraction
      of spikes in the radiative transfer simulations are due to
      hotter temperatures in recently ionized regions of the Universe.
   
    \item We have compared the three observed flux statistics with
      those derived from our optically thin \hot and \cold simulations
      in three redshift bins.  The  statistics for  the \hot and \default
      model are in significantly better agreement 
      with those from observations in all the 3 redshift bins. The 
        Doppler parameter distribution which is most sensitive to  the instantaneous 
        temperature of the gas  is in significantly better agreement for the \default model 
        than for both the \cold and \hot model.
      We constrain thermal parameters
      by varying $T_0$ and $\gamma$ in post-processed optically thin
      simulations. The best fit values at $5.3 \leq z \leq 5.5$,
      $5.5 \leq z \leq 5.7$, $5.7 < z < 5.9$ are $T_0
      \sim$ $11000 \pm 1600$, $10500 \pm 2100$, $12000 \pm 2200$ K 
      and $\gamma \sim$ $1.20 \pm 0.18$,$1.28 \pm 0.19$, $1.05 \pm 0.22$ 
      respectively.  We do not find significant evolution in $T_0$ and 
      $\gamma$ over $5.3 < z < 5.9$.

     \item We have also compared the three statistics in
       physically motivated radiative transfer simulations with those
       from observations.  Unlike optically thin simulations,
       radiative transfer simulations incorporate spatial fluctuations
       in the amplitude of the UVB and the TDR. As a result the
       scatter in the TDR is large and a single power-law cannot
       describe the TDR in radiative transfer simulations. The
       observed spike statistics in our radiative transfer simulation
       of late reionization with neutral islands persisting to $z \sim
       5.3$ are in good agreement {(see appendix \ref{app:statistics-comparison} 
       for details)} with observations in all three redshift bins .

\end{itemize}
 
Our work shows the potential of transmission spike shapes (heights and
widths) for constraining the thermal history of the
IGM near the tail-end of \HI reionization, complimentary to other
transmitted flux based methods.  In future, a much larger sample of
high resolution, high \SNR and high-$z$ QSO absorption spectra should
become available thanks to $30$--$40\rm\,m$ class optical
telescopes. These larger data sets, complemented by further improved
radiative transfer simulations, promise to put tight constraints on the
nature and the exact timing of \HI reionization.

%
%


\section*{Acknowledgment}

We thank the staff of the Las Campanas and Keck observatories for
their help with the observations.  MR thanks Ian Thompson and Steve
Shectman for suggestions for installing a new bandpass filter in the
MIKE slit-viewing camera.  Support by ERC Advanced Grant 320596 `The
Emergence of Structure During the Epoch of reionization' is gratefully
acknowledged. JSB acknowledges the support of a Royal Society
University Research Fellowship. GDB was supported by the National Science Foundation 
through grants AST-1615814 and AST-175140.
The \sherwood and \relics simulations
were performed with supercomputer time awarded by the Partnership for
Advanced Computing in Europe (PRACE) 8th and 16th calls.  We
acknowledge PRACE for awarding us access to the Curie and Irene
supercomputers, based in France at the Tr\'{e}s Grand Centre de Calcul
(TGCC).  This work also used the Cambridge Service for Data Driven
Discovery (CSD3), part of which is operated by the University of
Cambridge Research Computing on behalf of the STFC DiRAC HPC Facility
(www.dirac.ac.uk). The DiRAC component of CSD3 was funded by BEIS
capital funding via STFC capital grants ST/P002307/1 and ST/R002452/1
and STFC operations grant ST/R00689X/1. We also acknowledge the DiRAC
Data Intensive service at Leicester, operated by the University of
Leicester IT Services. The equipment was funded by BEIS capital funding via
STFC capital grants ST/K000373/1 and ST/R002363/1 and STFC DiRAC Operations
grant ST/R001014/1. We also thank DiRAC@Durham facility managed by the
Institute for Computational Cosmology on behalf of the STFC DiRAC HPC Facility
(www.dirac.ac.uk). The equipment was funded by BEIS capital funding via STFC
capital grants ST/P002293/1, ST/R002371/1 and ST/S002502/1, Durham University
and STFC operations grant ST/R000832/1. DiRAC is part of the National
e-Infrastructure.


\bibliographystyle{mnras}
\bibliography{spikes} 

\appendix


\section{Observational Systematics}
\label{app:observational-systematics}
\figref{fig:observed-spectra-example} shows \lya forest covered by 5 QSO
sightlines from our observed sample.  The observed spectra are subject to
systematics due to finite resolution, \SNR and continuum fitting uncertainties.
In this section, we quantify the effect of such systematics on spike statistics
and illustrate the method we use to account for these effects.

\subsection{Continuum fitting uncertainty}
\InputFigCombine{HIGH_RES_HIGH_SNR_DATA_WITH_PARAM_TICKS.pdf}{170}{
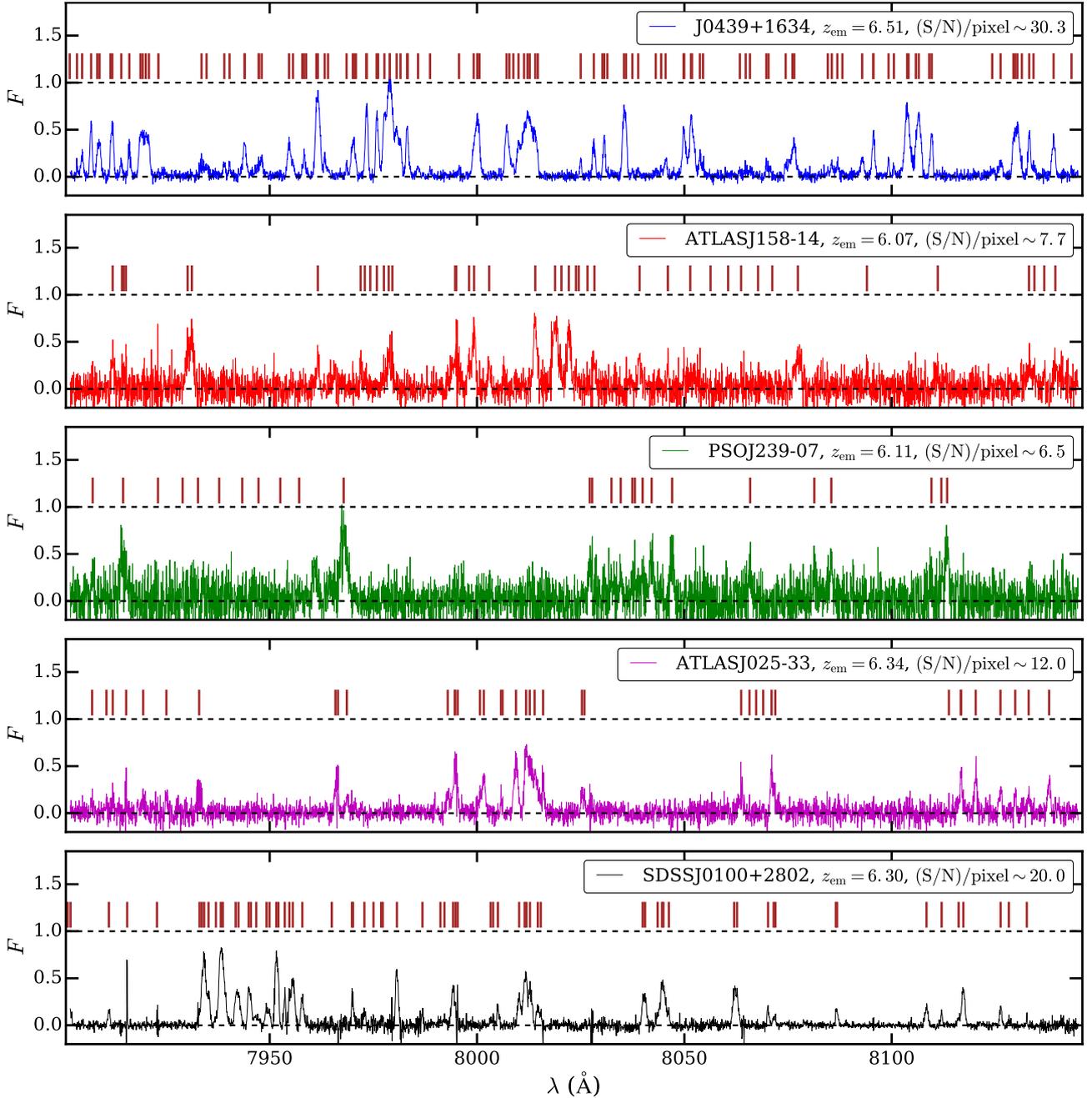
Examples of observed transmission spikes from the sample of 5 QSO sightlines.
The brown vertical ticks in each panel mark Voigt profile components 
identified and fitted using \viper.
}{\label{fig:observed-spectra-example}}

%

\InputFigCombine{Continuum_Uncertainty_Norm_Unnorm_BPDF_OD_Scaling_KDE.pdf}{175}{%
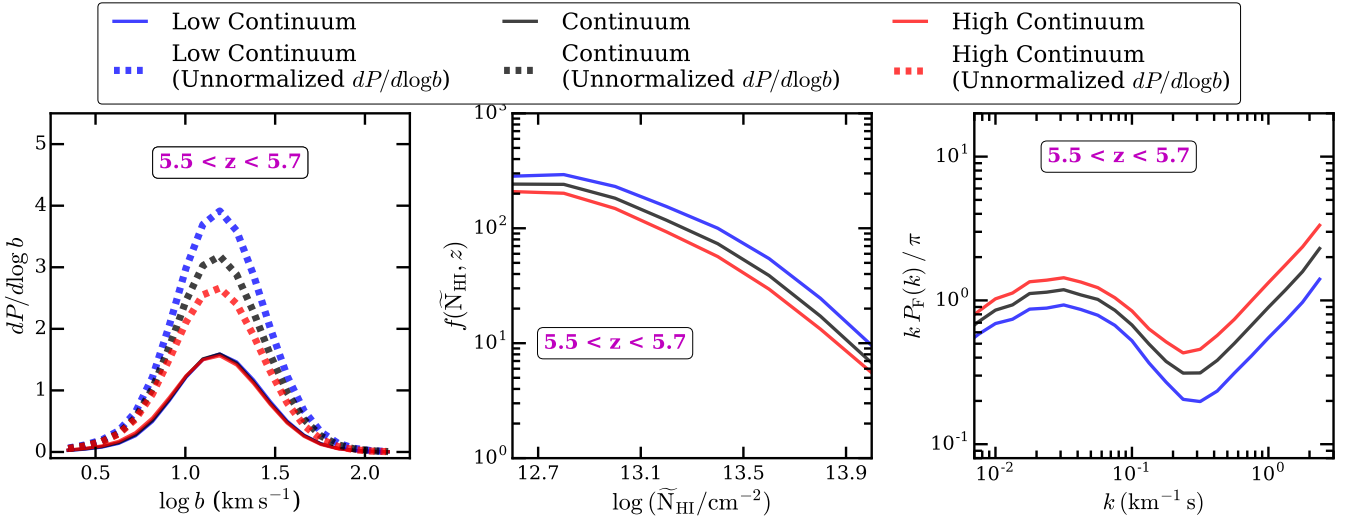
The effect of continuum fitting uncertainty on the three  statistics for
simulated spectra: the spike width distribution (left panel), pCDDF (middle
panel) and FPS (right panel) in \default model. The blue, black and red solid
curves represent the  statistics obtained using a low continuum ($F_{\rm cont}
- \delta F_{\rm cont}$, $\delta F_{\rm cont} \sim 0.25 F_{\rm cont}$), the
default continuum ($\delta F_{\rm cont}=0$) and a high continuum ($F_{\rm cont}
+ \delta F_{\rm cont}$, $\delta F_{\rm cont} \sim 0.25 F_{\rm cont}$)
respectively.  The normalization of the FPS and pCDDF are rather sensitive to
the continuum placement.  The continuum placement has remarkable little effect
on the normalized spike width distribution.  The unnormalized spike width
distribution (dotted lines in left panel) are significantly different due to
the difference in the number of fitted components (For visual purpose, the
spike width distribution and pCDDF are estimated using Gaussian kernel density
estimation. Note that the number of lines is scaled down by a factor of $\sim
8$ in the case of unnormalized spike width distribution.).  The mean flux
inferred for  the default continuum model is matched to the observed mean flux
by rescaling the optical depth to account for the  uncertainty in \GHI at $5.5
\leq z \leq 5.7$.  Note that  the mean flux  for the three continuum models
shown above is different.  In this work, we only use the normalized spike width
distribution to constrain $T_0$ and $\gamma$ as it is much less sensitive to
the continuum placement uncertainty than the two other statistics.
}{\label{fig:continuum-fit-spike-statistics-simulation}}
\InputFigCombine{Continuum_Fit_Example.pdf}{175}{
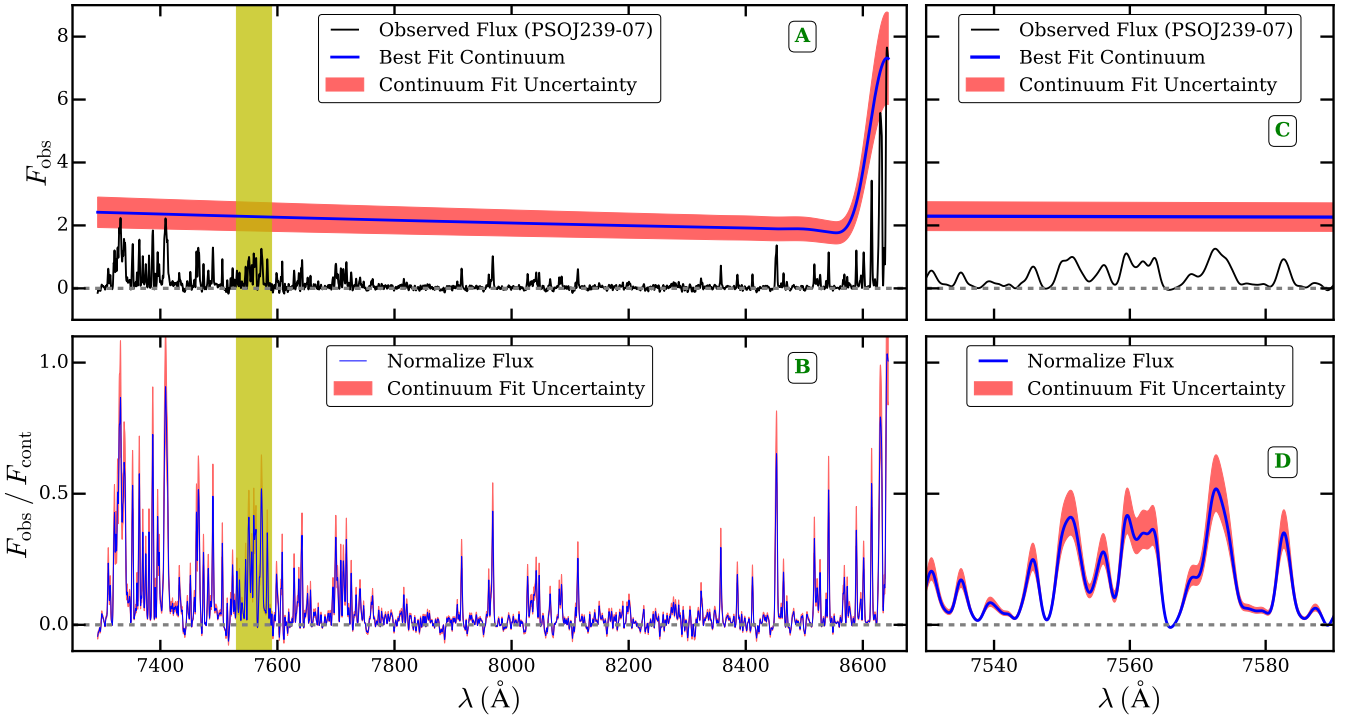
Panel A shows the observed flux ($F_{\rm obs}$) towards QSO PSOJ239-07
(black curve).  The blue line and red shaded region show the best fit
continuum ($F_{\rm cont}$) and the associated $1\sigma$ uncertainty,
respectively.  Panel B shows the normalized flux (blue line) obtained
by $F_{\rm norm} = F_{\rm obs} / F_{\rm cont}$.  For better visual
appearance of this figure, the flux is smoothed using a Gaussian
filter to reduce noise (we have not applied this Gaussian filter
anywhere else). Panels C and D are the same as panels A and B,
respectively, except that they show a zoomed version of the yellow
shaded region in panels A and B. We use these high and low continuum
fits to estimate the effect of continuum fitting uncertainty on
transmission spike properties.}{\label{fig:continuum-fit-example}}
\InputFigCombine{Effect_of_Continuum_Uncertainty.pdf}{175}{%
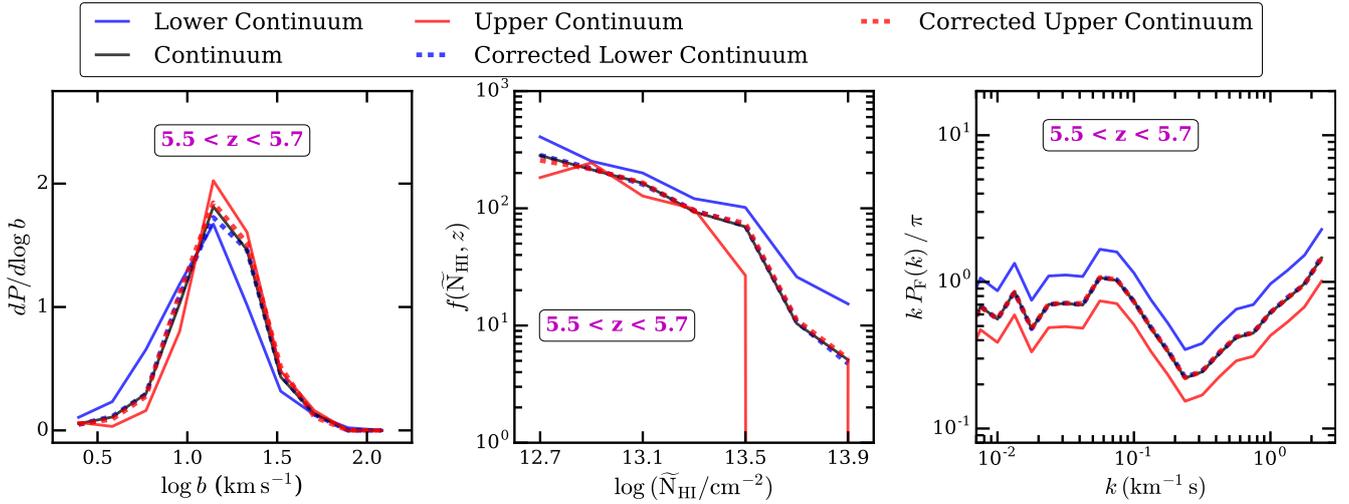
The effect of continuum fitting uncertainty on the three
statistics: the spike width distribution (left panel), pCDDF (middle
panel) and FPS (right panel). The blue, black and red solid curves
represent the observed spike statistics obtained using a low continuum
($F_{\rm cont} - \delta F_{\rm cont}$), the best fit continuum ($F_{\rm
cont}$) and a high continuum ($F_{\rm cont} + \delta F_{\rm cont}$)
respectively.  The normalization of the FPS and pCDDF are sensitive to the
observed continuum placement.  The continuum placement does not have a 
strong effect on the peak of the spike width distribution.   We  also show 
the effect of rescaling  the inferred optical depth  in the low and high continuum
models to match  mean flux inferred from the best fit continuum model.  The
``corrected'' low and high spike statistics are shown by blue and red
dotted lines, respectively. The distribution for the corrected low and high 
continuum model  are in good agreement with the distribution
corresponding to the best fit continuum.
}{\label{fig:continuum-fit-spike-statistics-observation}}

Due to the large opacities towards high redshift QSOs, the continuum
placement is non-trivial. We explained the method of our continuum fitting
in \S \ref{sec:spike-qualitative-analysis}. 
The matching of the  mean flux in observed and simulated spectra in a given
redshift bin is affected  by both the uncertainty in \GHI and the uncertainty
in the continuum placement of the QSO.  Unfortunately, there is significant
uncertainty in both \GHI and continuum placement of QSOs at high redshift.
Note however, that the scaling in flux is different for the two effects. While
continuum errors affect the inferred flux level linearly,  errors in the \GHI
assumed in the simulated spectra affect the optical depth linearly, but the
flux non-linearly due to the non-linear dependence of flux on optical depth ($F
= {\rm e}^{-\tau}$).  Matching the  observed mean flux by rescaling optical
depth (\GHI scaling) or rescaling flux (continuum uncertainty) have thus
different effects on the transmission spikes.  We illustrate the effect of flux
scaling (continuum fitting uncertainty) on the spike statistics in
\figref{fig:continuum-fit-spike-statistics-simulation} for the \default
simulation
model.  We calculate the three statistics by using a low continuum ($F_{\rm
cont} - \delta F{\rm cont}$), a default continuum ($\delta F_{\rm cont} = 0$) and a high
continuum ($F_{\rm cont} + \delta F{\rm cont}$).  The normalization of the FPS
and pCDDF is sensitive to the continuum placement.  Somewhat surprisingly the 
continuum placement does, however, have very little effect  on the spike width distribution. 
Note that the mean flux is different for the three models.
We have looked in some detail into individual simulated spectra and found that 
for a lower  placement of the continuum leading to  a linear  increase in observed flux levels 
the number of Voigt profile components increases. This appears to almost
perfectly compensate for the expected increase in the width of individual
components if the number of components stayed fixed for increased flux
levels.   Similarly, the effect of rescaling the optical depth (or \GHI) is
also small  for the spike width distribution, but significant for pCDDF and FPS
statistics  (see Appendix \ref{app:spike-statistics-sensitivity} and 
\figref{fig:param-effect-Gamma-12} for details).

We illustrate the effect of continuum fitting uncertainty on the shape of
 transmission spikes in our observed spectra in \figref{fig:continuum-fit-example}.  As
already discussed the  continuum
placement uncertainty (shown by red shaded region in panel A and C), leads us
to expect significant variation in the height of the spikes (panel D).
However, the width distribution of the Voigt profile components of the fits to the 
transmission spikes  is remarkably  robust.  To quantify the effect
of the continuum on  the observed transmission spikes, we show  the three statistics in
\figref{fig:continuum-fit-spike-statistics-observation} by using again a low
continuum ($F_{\rm cont} - \delta F{\rm cont}$), a best fit continuum ($F_{\rm
cont}$) and a high continuum ($F_{\rm cont} + \delta F{\rm cont}$).  Similar to 
\figref{fig:continuum-fit-spike-statistics-simulation}, the
normalization of the FPS and pCDDF in
\figref{fig:continuum-fit-spike-statistics-observation} is sensitive to the
continuum placement.  The continuum placement slightly changes the location of 
the peak in the spike width distribution and the distribution is somewhat
broader for a low continuum. Note that the observed distributions are less 
well defined than those from our simulated spectra   due to the small total number of spikes in
our observed spectra   compared to the simulated spectra\footnote{In
\figref{fig:continuum-fit-spike-statistics-simulation}, we calculated the
statistics from $5 \times 80=400$ simulated spectra.}.
We further emphasise again, that when comparing observed and simulated spectra 
we match the observed mean  flux so there is  a similar effect on the distributions of the 
simulated spectra.
To demonstrate the expected effect of matching the observed mean flux, we have
rescaled the optical depths  inferred from the observed spectra  (non-linear
flux scaling) in the high and low continuum model such that the mean flux
matches that with default continuum model. We call these continuum models
``corrected'' low or high continuum
models (as shown by dotted lines in
\figref{fig:continuum-fit-spike-statistics-observation}).  
The distributions for the corrected low
and high continuum models  are in very good agreement with that for 
the best fit continuum.  This demonstrates that the
effect of continuum fitting on spike statistics can indeed be minimized by
rescaling the simulated optical depth to match the mean observed flux as discussed earlier
for the simulated spectra.

\subsection{Effect of \SNR}
\InputFig{Sensitivity_Curve.pdf}{80}{%
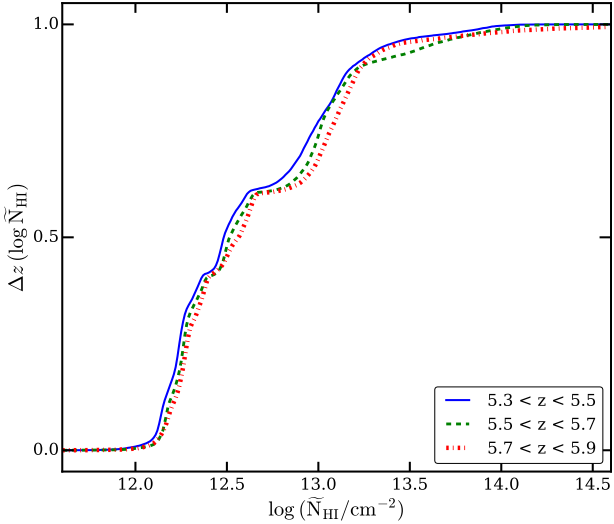
Calculation of sensitivity curve from observational sample in the
three different redshift bins. The sensitivity curve is calculated by
summing the total redshift path length in the observed spectra which
lies below the limiting equivalent width. The limiting equivalent
width is a theoretically expected equivalent width calculated from the
curve of growth. The sensitivity curve is shown for a $3 \sigma$
detection level.  The area under the sensitivity curve is used to
calculate the \HI pseudo-column density distribution
function.}{\label{fig:sensitivity-curve}}

In this section we illustrate the effect of finite \SNR on the
detectability of spikes.  Since high-$z$ QSOs are usually faint and
most of the observed pixels are close to $F \sim 0$, the noise in
observed spectra is mostly determined by the sky
background. Furthermore, the noise can vary along the wavelength
axis. To minimize the effect of finite \SNR on spike statistics, we
degrade the simulated spectra with noise generated from the
observed \SNR per pixel array. The statistics computed from observed
and simulated transmitted flux are consistent with each
other. However, the ability of the Voigt fitting procedure crucially
depends on the \SNR of the spectra. To account for this, we compute a
significance level (SL) for each Voigt component that accounts for
the \SNR, pixel separation and resolution of the
instrument \citep{gaikwad2017b}. We select the Voigt components with
SL $>3$. 

The finite \SNR of the observed spectra also sets the completeness
limit of the sample. However, one needs to account for the
incompleteness of the sample for the lines with \logaNHI below the
completeness limit.  We account for the incompleteness of the sample
by calculating the sensitivity curve as shown
in \figref{fig:sensitivity-curve}.  The plot shows the sensitivity
curve for three redshift bins. The observed sample is 50 per cent
complete for \logaNHI$\sim 12.5$. We use the area under the curve to
calculate the pCDDF and thus account for the effect of finite \SNR.

%
\section{Sensitivity of spike statistics to astrophysical parameters}
\label{app:spike-statistics-sensitivity}

\figref{fig:param-effect-T0} to \figref{fig:param-effect-L-jeans} shows
the sensitivity of our chosen statistics to the normalization of the
TDR ($T_0$), the slope of TDR ($\gamma$), the \HI photo-ionization
rate (\GHI) and Jeans smoothing, respectively.  We vary $T_0$, $\gamma$
and \GHI in the post-processing step by using a power-law TDR
$T=T_0 \Delta^{\gamma-1}$ and rescaling the optical depth under the
assumption of photo-ionization equilibrium. This approach allows us to
study the variation of spike statistics for a given parameter while
keeping other parameters fixed.

Fig.~\ref{fig:param-effect-T0} shows that all three transmission spike
statistics are sensitive to $T_0$. The spike width distribution (left
panel in \figref{fig:param-effect-T0}) is most sensitive to $T_0$
i.e., the spike width distribution is systematically shifted to larger
$b$ values for hotter models. The shape of the pCDDF (middle panel
in \figref{fig:param-effect-T0}) is also sensitive to $T_0$. This can
be understood by examining \figref{fig:los-comparison}, where we see
that spikes in hotter models are usually more blended than cold models
due to line of sight temperature and density smoothing effects.  Due
to such blending, \viper fits fewer components with low \logaNHI and
more components with high \logaNHI in hotter models.  The FPS (right
panel) is systematically lower at $0.1 < k \: ({\rm km^{-1} \: s}) <
1$ for $T_0=25000$ K compared to $T_0=10000$ K. This is expected as
the increase in temperature smoothes the transmitted \lya flux,
reducing the small scale power.

We illustrate the sensitivity of the three statistics to $\gamma$ in
\figref{fig:param-effect-gamma}.  It is not possible to vary the slope
of the TDR around the mean density without then varying the
temperature at the density where the spikes are most sensitive. For
example, the spikes in the optically thin simulations are most
sensitive to $\Delta \sim 0.3$
(See \figref{fig:delta-NHI-corr-cold-hot-aton-patchy}).  If we vary
$\gamma$ with the normalization of the TDR pivoted at $\Delta=1$,
(i.e. the $T_0$ value), we obtain a large variation in temperature at
$\Delta=0.3$. As a result it is hard to disentangle the effect of
variation in $T_0$ from $\gamma$.  To circumvent this problem, we
pivot the TDR at the densities where spikes are most sensitive i.e.,
$\Delta = 0.3$. Thus we use a power-law TDR of the form $T=T_{0.3} \:
[\Delta/0.3]^{\gamma-1}$.  The left and right panels in
\figref{fig:param-effect-gamma} show that the spike width distribution and FPS
are less sensitive to the variation in $\gamma$.  We see a slight
variation in the shape of the pCDDF for a variation in $\gamma$, such
that high \logaNHI systems are less frequent for $\gamma = 1.6$ than
for $\gamma=1$.  This is a direct consequence of the lower temperature
at $\Delta < 0.3$, since the spikes are more sensitive to $\Delta <
0.3$ than $\Delta > 0.3$.

The effect of the photo-ionization rate \GHI is illustrated
in \figref{fig:param-effect-Gamma-12}. The heights and number of
spikes (for a given \SNR) are sensitive to \GHI. As a result, the
normalization of the pCDDF is mostly sensitive to \GHI while the shape
remains relatively unchanged.  The normalization of the FPS depends on
the mean transmitted flux, and since the mean transmitted flux varies
with \GHI the normalization of the FPS is also different.  However,
the spike width distribution is relatively robust to large variations
in \GHI, allowing one to minimize the degeneracy between \GHI and
thermal parameters.  Note the effect of continuum placement
uncertainty is very similar to the variation in \GHI
(see \S \ref{app:observational-systematics}).

We show the recovery of $T_0$ and $\gamma$ for fiducial \hot and \cold model in
\figref{fig:T0-gamma-recovery}. We vary TDR using $\Delta$ field from \default
model for a range of $T_0$ and $\gamma$.  We use BPDF statistics to recover the
$T_0$ and $\gamma$ from \hot and \cold model.  \figref{fig:T0-gamma-recovery}
demonstrate that we can recover $T_0$ and $\gamma$ within $1\sigma$ for wide
range of $T_0$ and $\gamma$.

We study the effect of pressure (or Jeans) smoothing
in \figref{fig:param-effect-L-jeans}.  We take the density and
velocity fields from the \default, \hot and \cold optically thin
models and rescale the instantaneous temperatures and neutral hydrogen
fractions to have the same values.  Differences in the spike
statistics due to pressure smoothing are therefore isolated from the
effect of thermal broadening.  In all three models reionization
completes at $z\simeq 6.2$, following the UVB synthesis model
of \citet{puchwein2019}, with a total energy per proton mass of
$u_{0}=6.4 \rm\,eV\,m^{-1}_{\rm p}$ (\default), $u_{0}=12.4 \rm\,eV\,m^{-1}_{\rm
p}$ (\hot) and $u_{0}=3.4 \rm\,eV\,m^{-1}_{\rm p}$ (\cold) deposited over
the redshift interval $6<z<13$ \citep[cf.][]{nasir2016}. For
comparison, \citet{boera2019} have recently inferred
$u_{0}=4.6^{+1.4}_{-1.2}\rm\,eV\,m^{-1}_{\rm p}$ over the redshift interval
$6<z<13$ from new measurements of the \lya FPS at $z=5$, which is
consistent with our \default simulation at $\sim 1.3\sigma$.
Interestingly, \figref{fig:param-effect-L-jeans} shows that the spike
width distribution is only modestly sensitive to the pressure
smoothing, despite the different integrated thermal histories in the
models.  This is in part because, relative to the IGM at $z\leq 5$,
gas has had slightly less time to dynamically respond to changes in
the pressure following the completion of reionization at $z=6.2$, and
in part because the transmission spikes become rapidly more sensitive
to the most highly underdense gas as redshift increases (where the
dynamical time scales as $t_{\rm dyn}=\sqrt{\pi/G\rho}\simeq
H(z)^{-1}\Delta^{-1/2}$).  We estimate that systematic uncertainties
in the Jeans smoothing will therefore impact on the recovery of
$T_{0}$ by at most $\delta T_{0} \sim 600\rm\,K$ (see
Fig.~\ref{fig:T0-gamma-constraints-z-reion}).  We add this
uncertainty in quadrature to the final measurements we present for
$T_{0}$.  We have furthermore verified if we take a more extreme
model where the IGM is ionized rapidly around $z=15$, (i.e. the
fiducial model in the original \sherwood simulation suite), larger
differences in the spike width distribution due to Jeans smoothing are
present.  We argue here, however, that such an early end to
reionization is unlikely.

\InputFigCombine{L40N2048_COLD_T0_Variation.pdf}{175}{
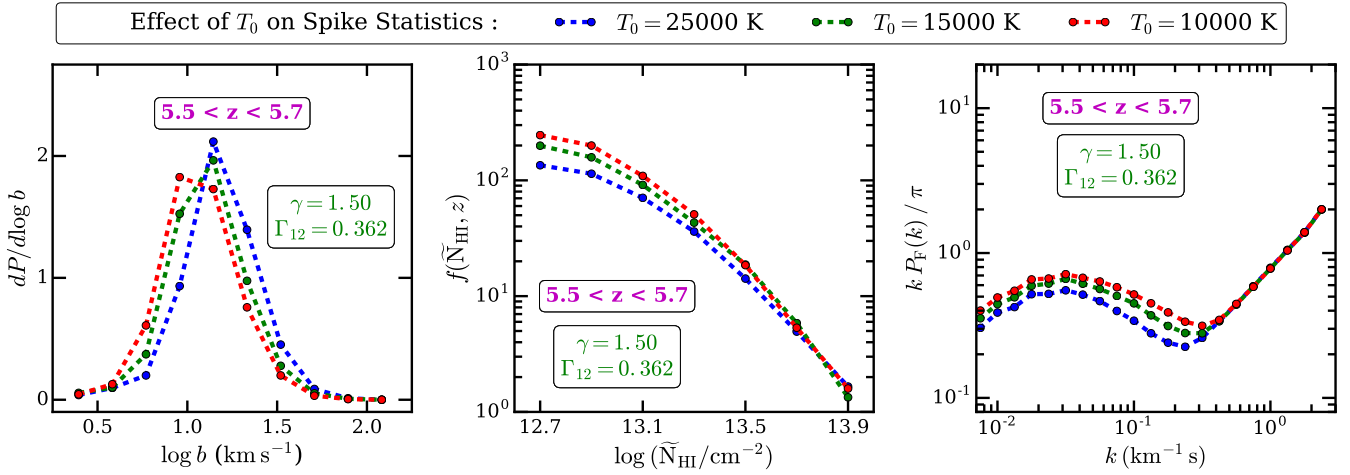
The effect of varying $T_0$ on spike statistics, showing the spike
width distribution (left panel), pCDDF (middle panel) and FPS (right
panel).}{\label{fig:param-effect-T0}}

\InputFigCombine{L40N2048_COLD_gamma_pivot_Variation.pdf}{175}{
As for \figref{fig:param-effect-T0}, except the effect of variation in
$\gamma$ on the spike statistics is
illustrated.}{\label{fig:param-effect-gamma}}


\InputFigCombine{L40N2048_COLD_Gamma_12_Variation.pdf}{175}{%
As \figref{fig:param-effect-T0} except the effect of variation in \GTW
on the spike statistics is
illustrated.}{\label{fig:param-effect-Gamma-12}}

\InputFigCombine{T0_gamma_Recovery.pdf}{175}{%
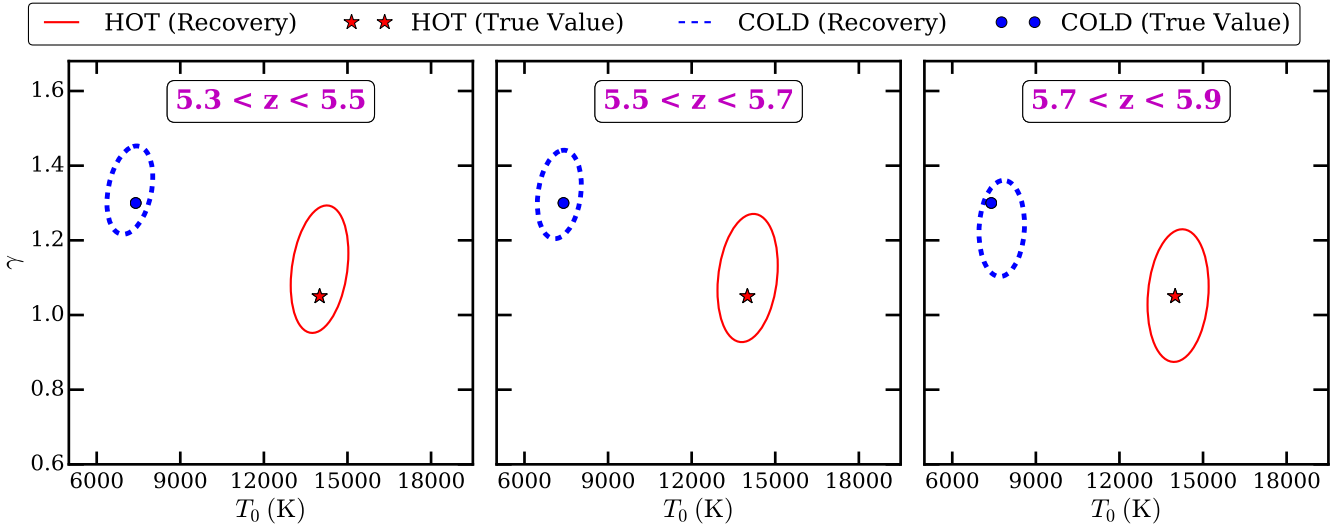
As for panel A in \figref{fig:T0-gamma-constraints}, except the recovery of
$T_0$ and $\gamma$ from \hot and \cold model is illustrated. The TDR is varied
using $\Delta$ field from \default model for a given value of $T_0$ and $\gamma$.
We use BPDF statistics to recover the $T_0$ and $\gamma$ from \hot and \cold model.
We can recover $T_0$ and $\gamma$ within $1\sigma$ for wide range of $T_0$ and $\gamma$.
}{\label{fig:T0-gamma-recovery}}

\InputFigCombine{L40N2048_Jeans_Smoothing.pdf}{175}{
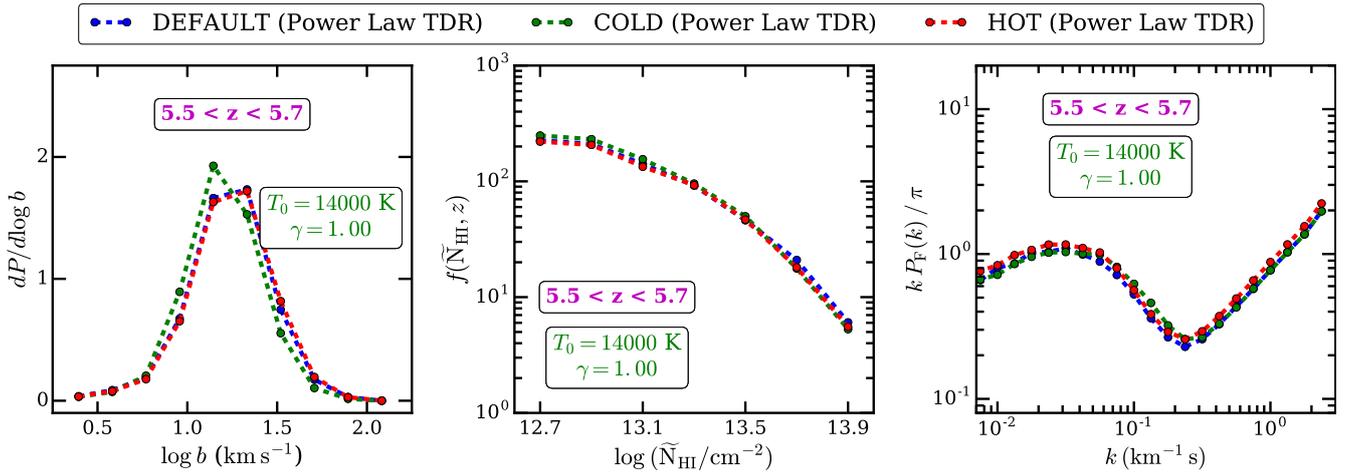
As for \figref{fig:param-effect-T0} except the effect of variation in
the pressure smoothing scale on the spike statistics is
illustrated.}{\label{fig:param-effect-L-jeans}}

\InputFigCombine{T0_gamma_constraints_jeans_smoothing_effect.pdf}{175}{%
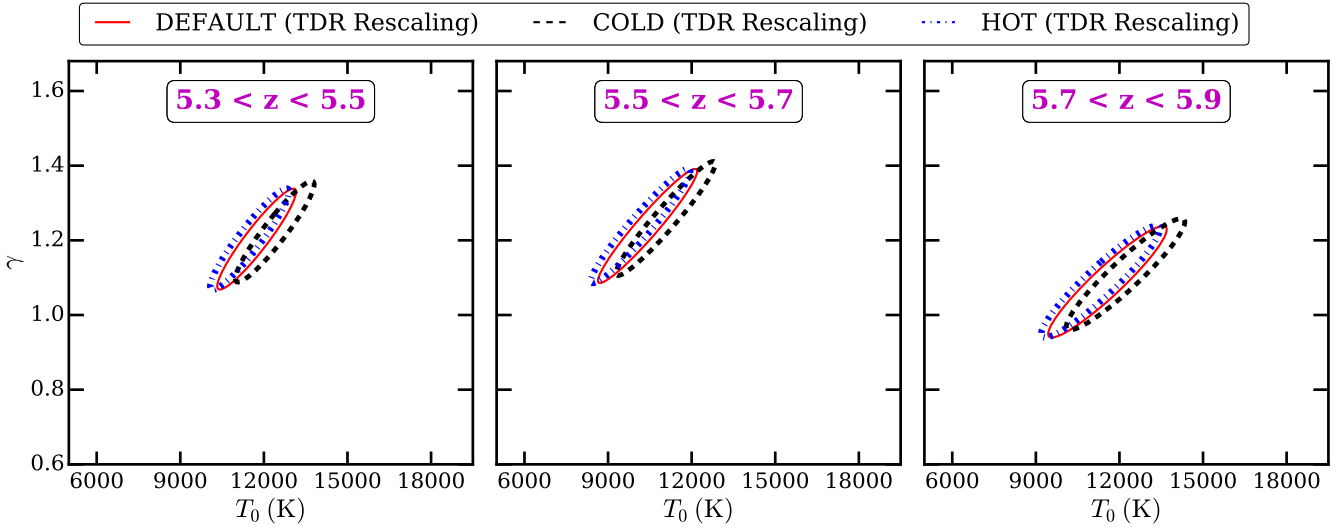
As for panel A in \figref{fig:T0-gamma-constraints}, except the effect
of variation in the pressure smoothing scale on the $T_0-\gamma$
constraints is quantified in 3 redshift bins. }{\label{fig:T0-gamma-constraints-z-reion}}

%

\section{Numerical effects}
\label{app:resolution-study}
\InputFigCombine{{Resolution_Effect_z-5.6_Box_Size}.pdf}{175}{
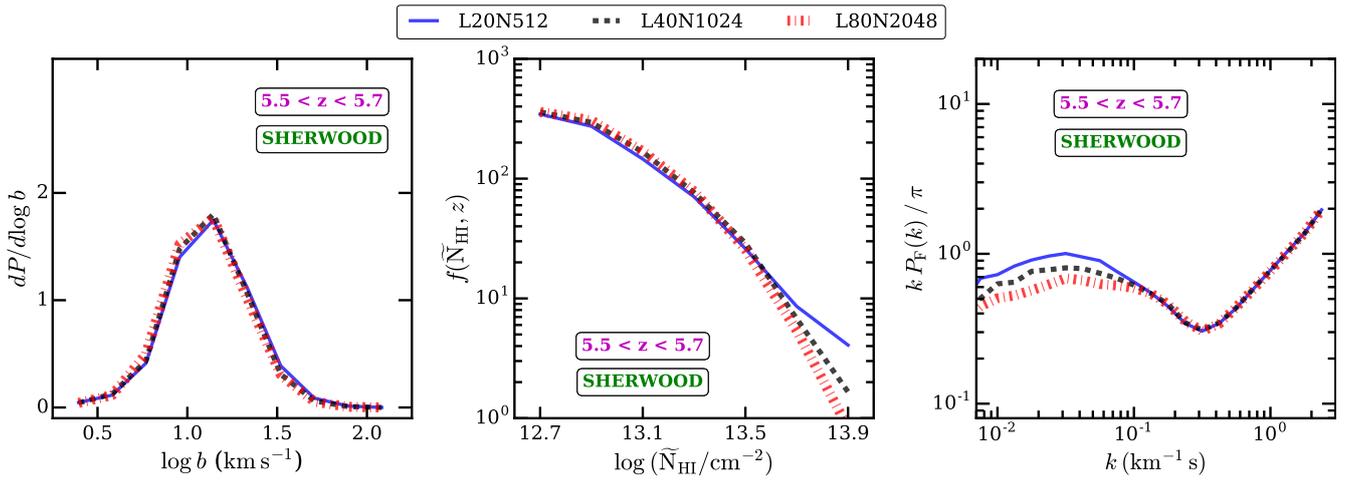
The effect of box size on the spike width distribution (left panel),
pCDDF (middle panel) and FPS (right panel) in the \sherwood simulation
suite at $z=5.6$, for box sizes of $20h^{-1}\rm\,cMpc$ (L20N512),
$40h^{-1}\rm\,cMpc$ (L40N1024) and $80h^{-1}\rm\,cMpc$ (L80N2048) at
fixed mass resolution, $m_{\rm gas} \sim 7.97 \times 10^5 \: {\rm
M_{\odot}}$.  The results are relatively well converged for the spike width
distribution.
}{\label{fig:resolution-effect-1}}

\InputFigCombine{{Resolution_Effect_z-5.6_Mass_Resolution}.pdf}{175}{
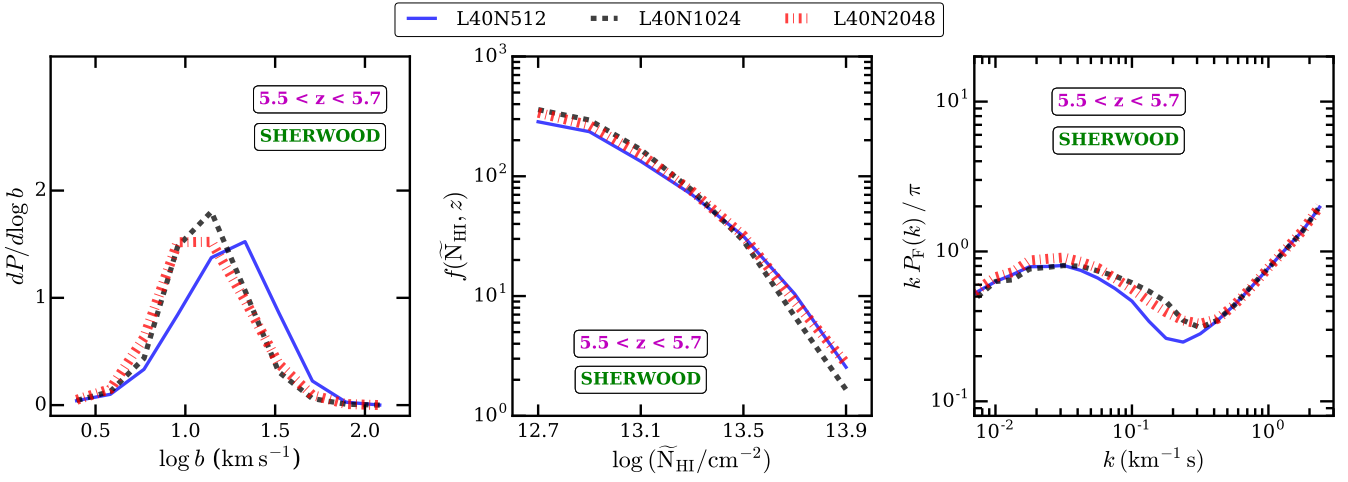
The effect of mass resolution on the spike width distribution (left
panel), pCDDF (middle panel) and FPS (right panel) in the \sherwood
simulation suite at $z=5.6$, for a gas particle mass of $m_{\rm
gas} \sim 6.38 \times 10^6 \: {\rm M_{\odot}}$ (L40N512), $m_{\rm
gas} \sim 7.97 \times 10^5 \: {\rm M_{\odot}}$ (L40N1024) and $m_{\rm
gas} \sim 9.97 \times 10^4 \: {\rm M_{\odot}}$ (L40N2048) for a fixed
box size of $40h^{-1}\rm\,cMpc$.  High mass resolution is important
for correctly resolving the widths of the transmission spikes.}
{\label{fig:resolution-effect-2}}

We assess the effect of simulation box size and mass resolution on
convergence of the spike statistics
in \figref{fig:resolution-effect-1}
and \figref{fig:resolution-effect-2}. We use simulations with varying
box sizes and gas particle masses drawn from the optically
thin \sherwood simulation suite
\citep{bolton2017}.  All other parameters such as cosmology and the UVB evolution are
same for the simulations.  The spike statistics are well converged for
our fiducial box size and mass resolution, which corresponds to the
L40N2048 model.

\figref{fig:chi-square-field-comparison} tests our method of constraining
$T_0-\gamma$ from the spike width distribution using a $\chi^2$
minimization process by evaluating the likelihood $\mathscr{L} = {\rm
e}^{-\chi^2/2}$.  The best fit model corresponds to a minimum $\chi^2$
$(\chi^2_{\rm min})$, where our $1\sigma$ constraints on $T_0-\gamma$
are contours of constant $\chi^2 = \chi^2_{\rm min} + \Delta \chi^2$
where we assume $\Delta \chi^2 = 2.30 $ for 2 degrees of
freedom \citep{avni1976,press1992}.  To a good approximation, we
confirm the $\chi^2$ distribution in $T_0-\gamma$ plane is Gaussian.

\InputFig{Chi_Square_Field_Comparison.pdf}{80}{%
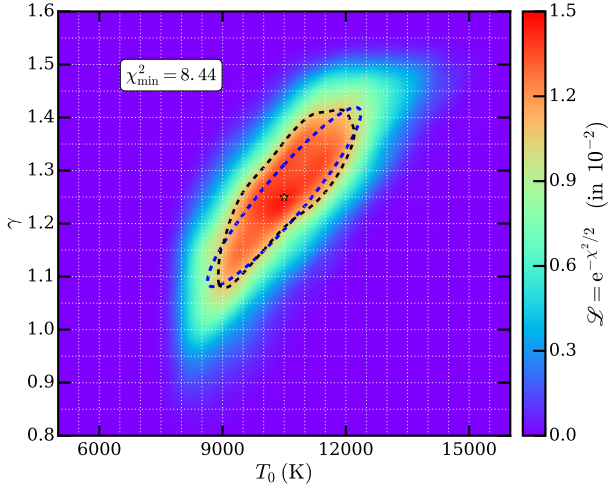
The likelihood $\mathscr{L} = {\rm e}^{-\chi^2/2}$ obtained by
comparing the simulated spike width distribution to observations at
$5.3 \leq z \leq 5.5$.  The best fit model corresponds to model with
$\chi^2_{\rm min}=8.44$ (yellow star).  The black dashed line shows
the contours of $\chi^2_{\rm min} + \Delta \chi^2 = 10.7$ where
$\Delta \chi^2 = 2.30 $ for 2 degrees of freedom ($\chi^2_{dof}
\sim 1.05$).  The blue dashed line shows $1\sigma$ constraints on $T_0-\gamma$
assuming $\chi^2$ (and hence $\mathscr{L}$) is Gaussian distributed.
To a good approximation, the $\chi^2$ distribution in the $T_0-\gamma$
plane is Gaussian.  The grid shows the sampling of points in
$T_0-\gamma$ plane used to obtain the $\chi^2$ field.}
{\label{fig:chi-square-field-comparison}}

\section{Detailed comparison of statistics from observations with models}
\label{app:statistics-comparison}
    In this section, we first discuss quantitatively how well  the spike width
    distribution, pCDDF and FPS statistics from optically thin and radiative
    transfer simulations fit  the  observed distributions. \tabref{tab:chi-sq}
    shows the goodness of fit (reduced $\chi^2$) between model and observed
    statistics in the three redshift bins.  The goodness of fit between
    observed spike width distribution for the  \default model is significantly
    better than that for the \hot and \cold models in all 3 redshift bins. The
    pCDDF and FPS statistics of the  \hot model are in somewhat better
    agreement with observations than the corresponding statistics of the
    \default and \cold models. Overall the \default and \hot models therefore
    show a better fit  than the \cold model.  All  three statistics for the
    \aton model show a smaller  $\chi^2_{\rm dof}$ than the \patchy models, but
    the fits are generally reasonably good for both models. An exception to
    this is noticeable in the  $z=5.6$ redshift bin, where the spike width
    distribution and FPS are marginally better fit by the \patchy than by the
    \aton model. When we compare the goodness of fit between optically thin and
    radiative transfer simulations we find the following. Formally the spike
    width distribution in the \aton model is a better fit to observations than
    the \default model (except at $z=5.6$).  The $\chi^2_{\rm dof}$ of FPS and
    pCDDF statistics for the  \hot models are overall comparable to that from
    the \aton model. However, FPS and pCDDF statistics are also sensitive to
    continuum placement uncertainty as shown in appendix
    \ref{app:observational-systematics}. Hence in this work, we have focused on
    the spike width distribution for constraining $T_0$ and $\gamma$. As
    discussed in detail in the main text we consider the radiative transfer
    models \aton and \patchy as  more physical than the \default and \hot model
    because they account for spatial fluctuations of  the UVB and TDR.  As
    discussed in the introduction such fluctuations are important for producing
the observed \taueffHI scatter and long absorption troughs at these redshifts.

\begin{table*}
\centering
\caption{Reduced $\chi^2$ (i.e., $\chi^2$ per degree of freedom) between model and observed statistics}
\begin{threeparttable}
\begin{tabular}{cccccccccc}
\hline \hline
& \multicolumn{3}{c}{$z=5.4$} \hspace{10mm} & \multicolumn{3}{c}{$z=5.6$} \hspace{10mm} & \multicolumn{3}{c}{$z=5.8$} \vspace{2mm}\\ 
Simulation \hspace{5mm} & $b$ PDF & pCDDF & FPS \hspace{10mm} & $b$ PDF & pCDDF & FPS \hspace{10mm} & $b$ PDF & pCDDF & FPS \\ 
\hline
\default \hspace{5mm} & 2.27  & 4.01  & 1.00\hspace{10mm} & 1.10  & 0.68 & 0.84\hspace{10mm} & 0.88  & 0.49 & 0.52 \\
\cold    \hspace{5mm} & 14.69 & 15.24 & 6.69\hspace{10mm} & 10.08 & 2.82 & 0.79\hspace{10mm} & 11.40 & 1.47 & 0.89 \\
\hot     \hspace{5mm} & 5.10  & 1.60  & 0.62\hspace{10mm} & 5.26  & 0.57 & 1.34\hspace{10mm} & 1.12  & 0.31 & 0.36 \\
\aton    \hspace{5mm} & 1.91  & 2.06  & 0.63\hspace{10mm} & 1.51  & 0.55 & 1.66\hspace{10mm} & 0.49  & 0.35 & 0.24 \\
\patchy  \hspace{5mm} & 3.60  & 3.11  & 1.52\hspace{10mm} & 1.17  & 0.74 & 1.59\hspace{10mm} & 1.44  & 0.92 & 0.39 \\
\hline \hline
\end{tabular}
\end{threeparttable}
\label{tab:chi-sq}
\end{table*}






\bsp	
\label{lastpage}
\end{document}